\documentclass{aa}  

\usepackage{graphicx}
\usepackage{txfonts}
\usepackage{lineno}
\usepackage{subfigure}
\usepackage{booktabs}
\usepackage{xcolor} 
\usepackage{amsmath}

\newcommand\kms{{\rm\,km\,s^{-1}}}
\newcommand\msun{\rm\,M_\odot}

\newcommand\mso{$\,\mathrm{M}_\odot\,$}

 \def\simle{\mathrel{\hbox{\rlap{\hbox{\lower4pt\hbox{$\sim$}}}\hbox{$<$}}}}
 \def\simgr{\mathrel{\hbox{\rlap{\hbox{\lower4pt\hbox{$\sim$}}}\hbox{$>$}}}}

\renewcommand{\thefigure}{\arabic{figure}}

\begin{document}

   \title{
   The initial spin distribution of B-type stars revealed by the
   split main sequences of young star clusters
   }

   \author{Chen Wang 
          \inst{1}
, Ben Hastings \inst{2}, Abel Schootemeijer \inst{2}, Norbert Langer \inst{2}, Selma E. de Mink \inst{1,3},
         Julia Bodensteiner \inst{4},  Antonino Milone \inst{5}, Stephen Justham \inst{1}
          \and
          Pablo Marchant \inst{6}
          }

  \institute{Max Planck Institute for Astrophysics, Karl-Schwarzschild-Strasse 1, 85748 Garching, Germany\\
              \email{cwang@mpa-garching.mpg.de}
              \thanks{}
              \and
Argelander-Institut f\"ur Astronomie, Universit\"at Bonn, Auf dem H\"ugel 71, 53121 Bonn, Germany 
\and
Astronomical Institute ``Anton Pannekoek", University of Amsterdam, Science Park 904, 1098 XH Amsterdam, The Netherlands
\and
European Southern Observatory, Karl-Schwarzschild-Strasse 2, 85748 Garching, Germany
\and
Dipartimento di Fisica e Astronomia ``Galileo Galilei'', Univ. di Padova, Vicolo dell'Osservatorio 3, I-35122 Padova, Italy
\and
Institute of astrophysics, KU Leuven, Celestijnlaan 200D, 3001 Leuven, Belgium         
             }

   \date{Preprint online version: August 02, 2022}
\titlerunning{The initial spin distribution of B-type stars}
\authorrunning{Wang et al.}
 
  \abstract
   {Spectroscopic observations of stars in young open clusters have revealed evidence for a dichotomous distribution of stellar rotational velocities, with 10$-$30\% of stars rotating slowly and the remaining 70$-$90\% rotating fairly rapidly.
At the same time, high-precision multiband photometry of young star clusters shows a split main sequence band, which is again interpreted as due to a spin dichotomy. Recent papers suggest that extreme rotation is required to retrieve the photometric split.
Our new grids of MESA models and the prevalent SYCLIST models show, however, that initial slow (0$-$35\% of 
the linear Keplerian rotation velocities) and intermediate (50$-$65\% of the Keplerian rotation velocities) rotation are adequate to explain the photometric split. These values are consistent with the recent spectroscopic measurements of cluster and field stars, and are likely to reflect the birth spin distributions of upper main-sequence stars. A fraction of the initially faster-rotating stars may be able to reach near-critical rotation at the end of their main-sequence evolution and produce Be stars in the turn-off region of young star clusters.
However, we find that the presence of Be stars up to two magnitudes below the cluster turnoff advocates for a crucial role of binary interaction in creating Be stars. We argue that surface chemical composition measurements may help distinguish these two Be star formation channels. While only the most rapidly rotating, and therefore nitrogen-enriched, single stars can evolve into Be stars,
slow pre-mass-transfer rotation and inefficient accretion allows for mild or no enrichment
even in critically rotating accretion-induced Be stars.
Our results shed new light on the origin of the spin distribution of young and evolved
B-type main sequence stars.
}


   \maketitle
%

\section{Introduction} \label{sec:intro}
Rotation has a profound impact on stellar structure and evolution \citep[see][for a review]{Maeder2000}. On the one hand, it makes a star oblate, reducing its effective gravity, thereby reducing its equatorial flux and temperature, as described by the von Zeipel theorem \citep{Vonzeipel1924}. This is the so-called gravity darkening. Since the centrifugal force is larger at the equator than on the pole, a rotating star appears dimmer and cooler if it is seen equator-on as opposed to pole-on.
On the other hand, rotation is also predicted to trigger internal mixing in the stellar interior \citep{1978ApJ...220..279E}, such that fresh hydrogen is injected into the stellar center, while nuclear-burning products are dragged up to the stellar surface. 
As a consequence, rotationally induced mixing may extend the main-sequence (MS) lifetime of a rotating star and, on average, may make it more luminous and hotter than its nonrotating counterpart.
The consequences of rotation on the position of a star in the Hertzsprung-Russell diagram (HRD) or color-magnitude diagram (CMD) are determined by these effects. 
Thereby, rotation may affect the age determinations of star clusters based on single star models. 

Spectroscopic observations in the Tarantula nebula revealed a genuinely bimodal rotational velocity distribution for the early B-type stars (7$-$15\mso), with one peak at $0\leq v_\mathrm{e} \leq 100\kms$ and the other at $200 \leq v_\mathrm{e} \leq 300\kms$ \citep{Dufton2013}, where $v_\mathrm{e}$ is the equatorial velocity. 
A similar bimodal rotational velocity distribution is also reported in the Galactic field for late B- and early A-type stars in a mass range of 2.5$-$4\mso \citep{Zorec2012}, whereas stars with masses between 1.6\mso and 2.3\mso are found to have a unimodal equatorial velocity distribution peaking at $v_\mathrm{e} \sim 200\kms$ \citep{Zorec2012}.

It is interesting that such a bimodal rotational velocity distribution appears to be required to explain the split MSs detected in young star clusters as well (younger than $\sim$700\,Myr) in the
Large and Small Magellanic Clouds (LMC and SMC, respectively) based on Hubble Space Telescope (HST) \citep{Bastian2009,Girardi2011,Yang2013,Brandt2015,Niederhofer2015,Antona2015,
Correnti2017,2017ApJ...844..119L,2022NatAs...6..480W}, with the blue MS containing 10--30\% of the stars and the red MS containing the remaining stars. Other mechanisms including a prolonged star formation history, different metallicities, or different chemical compositions fail at explaining the split MS feature \citep[see][and references therein]{Milone2016}. Although it has been widely accepted that the red and blue MSs are comprised of stars with fast and slow rotation, respectively,
the difference in rotation speed which is required to explain the observed color split of the two MSs is still under discussion
\citep{Milone2018,2019ApJ...887..199G,2022NatAs...6..480W}. Most of the papers explain the red and blue MSs as comprised of stars rotating at 90\% and 0\% of their critical velocities by comparing the prevalent SYnthetic CLusters Isochrones \& Stellar Tracks (SYCLIST) single-star models\footnote{ https://www.unige.ch/sciences/astro/evolution/en/database/syclist/} \citep{2013A&A...553A..24G} with the photometric observations \citep[see][and references therein]{Antona2017,Milone2018}.
Interpreting the red MS stars with extremely fast rotation is further supported by a large fraction of Be stars, that is \ the B-type stars with emission lines that probably arise from decretion disks that are induced by near-critical rotation, detected in the Magellanic Cloud clusters younger than $\sim$300\,Myr \citep{1998A&A...340..397K,Bastian2017,Milone2018}.

However, it is unlikely that the majority of stars are born with near-critical rotation.
\citet{Huang2010} argued that nonevolved critically rotating B-type stars should be scarce ($<$ 1.3\%).
Meanwhile, the peak of the fast component in the abovementioned bimodal velocity distribution of field B- and A-type stars \citep{Zorec2012,Dufton2013} only corresponds to $\sim$50--60\% of the critical velocities.
The most compelling piece of evidence against the red MS stars rotating near critically arises from the spectroscopic measurement of the velocity of the red MS stars in the young LMC cluster NGC\,1818, which shows a mean projected equatorial rotational velocity $v\sin i$ of $\sim 202 \pm 23 \kms$, again only corresponding to velocities slightly larger than $\sim$ 50\% of their critical values \citep{Marino2018}.
In fact, several studies have indeed argued that the red MS should contain stars with around half-critical velocities, by comparing the single-star models computed from the Modules for Experiments in Stellar Astrophysics (MESA) code with the photometric observations \citep{2019ApJ...887..199G,2022NatAs...6..480W}. 

In this paper, we show that, counter-intuitively, this half-critical rotational velocity is in agreement with previously inferred near-critical rotational velocity, and that the differences are the result of, firstly, how the fraction of critical rotation is defined and, secondly, an early spin-down phase of the SYCLIST models.
The disadvantage of \citet{2019ApJ...887..199G} is that they investigated the impact of rotation in cluster turn-off stars, in which the influence of single star evolution and binary interaction cannot be eliminated.
 \citet{2022NatAs...6..480W} used their MESA isochrones with different rotational velocities to fit cluster split MSs, but they neither compared them with other stellar models nor compared them with the recent spectroscopic velocity measurements of cluster stars.

We aim to explore the similarities and the differences of different model sets when using them to study young star clusters.
We derive the rotational velocity of the red and blue MS stars in young star clusters by comparing their observed color split with the rotationally induced color variation in stellar models. 
We then directly compare stellar models with the photometric and spectroscopic observations of the MS stars in the LMC cluster NGC\,1818. After finding out the differences in the SYCLIST and the MESA models, we propose a possible way to probe the related physics assumptions in these two models. At last, we examine how the derived rotational velocity of the red MS stars affects our understanding of the origin of Be stars.



The layout of the paper is as follows. Section\,\ref{sec:def} clarifies different definitions of critical rotational velocity used in different stellar models. Section\,\ref{sec:method} describes the physics and assumptions adopted in computing our single-star models. Our results and comparisons with the observations are presented in Section\,\ref{sec:Results}. We also discuss our results and their implication on the origin of Be stars in this section. At last, we summarize our conclusions in Section\,\ref{sec:conclusion}.

\section{Different definitions of critical rotational velocity}\label{sec:def}
There are different definitions of the critical rotational velocity \citep{2013A&ARv..21...69R}, which need to be clarified before using stellar models. The critical velocity considered in the SYCLIST models uses the equatorial radius of a critically rotating star. 
Under this definition, the linear rotational velocity is expressed as:
\begin{equation}
v_\mathrm{crit, SYCLIST} = \sqrt{\frac{GM}{R_\mathrm{e,crit}}}=\sqrt{\frac{2}{3}\frac{GM}{R_\mathrm{p,crit}}}.
\end{equation}
Here $R_\mathrm{e,crit}$ and $R_\mathrm{p,crit}$ are the equatorial radius and the polar radius of a critically rotating star, respectively. The factor $2/3$ comes from the relation $R_\mathrm{e,crit}=3/2R_\mathrm{p,crit}$ in a critically rotating star, in the framework of the Roche approximation. 
Critical velocity can also be expressed in terms of angular velocity, which in the SYCLIST models is
\begin{equation}
\Omega_\mathrm{crit, SYCLIST} = \sqrt{\frac{GM}{R^3_\mathrm{e,crit}}}=\sqrt{\frac{8}{27}\frac{GM}{R^3_\mathrm{p,crit}}}.
\end{equation}
In general, the variation of the polar radius due to rotation is marginal, such that $R_{\mathrm{p,crit}}/R_{\mathrm{P}} \simeq 1$ is a good approximation, where $R_\mathrm{P}$ is the polar radius of a star rotating at an arbitrary rate. Therefore, it is possible to calculate $\Omega_\mathrm{crit, SYCLIST}$ and $v_\mathrm{crit, SYCLIST}$ without having a numerical model of a critically rotating star. This definition takes into account the fact that a star will increase its radius as it approaches critical rotation.

However, the MESA stellar models, including our newly computed models and the MIST models, adopt a different definition, in which the current equatorial radius $R_{\mathrm{e}}$ is used. The critical linear and angular velocities are then defined as:
\begin{equation}
\label{eq:3}
v_\mathrm{crit, MESA} = \sqrt{\frac{GM}{R_\mathrm{e}}} \quad \mathrm{and} \quad \Omega_\mathrm{crit, MESA}=\sqrt{\frac{GM}{R^3_\mathrm{e}}},
\end{equation}
respectively, which are denoted by $v_\mathrm{orb}$ and $\Omega_\mathrm{orb}$ in \citet{2013A&ARv..21...69R}.
This definition indicates the orbital velocity at the equator of a rotating star. As has been stated in \citet{2013A&ARv..21...69R}, a star rotates critically when its equatorial velocity equals the orbital velocity at its equator. In the case of a critically rotating star, $v_\mathrm{crit,SYCLIST}=v_\mathrm{crit,MESA}$ due to the increase of its equatorial radius. But for noncritically rotating stars, the MESA definition is more meaningful when studying fast-rotating stars with disks, because $v_\mathrm{crit,MESA}$ describes which velocity is required for a rotating star to eject material.

The critical velocity fraction is more frequently used than the absolute value of rotational velocity when describing how quickly a star rotates. 
Under the SYCLIST definition, the critical linear velocity fraction does not equal the critical angular velocity fraction. The relation between them is given by: 
\begin{equation}
\begin{split}
\frac{v_\mathrm{e}}{v_{\mathrm{crit, SYCLIST}}} & =\frac{\Omega_\mathrm{e} R_{\mathrm{e}}}{\Omega_{\mathrm{crit, SYCLIST}} R_{\mathrm{e,crit}}}\\
&=\frac{\Omega_\mathrm{e}}{\Omega_{\mathrm{crit, SYCLIST}}}\frac{R_{\mathrm{e}}}{R_{\mathrm{p}}}\frac{R_{\mathrm{p}}}{R_{\mathrm{p,crit}}}\frac{R_{\mathrm{p,crit}}}{R_\mathrm{e,crit}} \\ & \simeq \frac{2}{3}\frac{R_\mathrm{e}}{R_\mathrm{p}}\frac{\Omega_\mathrm{e}}{\Omega_\mathrm{crit, SYCLIST}}.
\end{split}
\end{equation} 
Here $v_\mathrm{e}$ and $\Omega_\mathrm{e}$ are the equatorial linear and angular velocities, respectively. 
Under the MESA definition, the fractional linear and angular velocities are identical, that is $v_\mathrm{e}/v_\mathrm{crit, MESA}=\Omega_\mathrm{e}/\Omega_\mathrm{crit, MESA}$. Except for the extreme cases of zero rotation and critical rotation, we have $v_\mathrm{e}/v_\mathrm{crit, MESA}<v_\mathrm{e}/v_\mathrm{crit, SYCLIST}<\Omega_\mathrm{e}/\Omega_\mathrm{crit, SYCLIST}$.

In this work, we always mean the MESA definition when referring to critical rotation, that is $v_\mathrm{crit}=v_\mathrm{crit, MESA}$, because we mainly use our MESA models to investigate the role of rotation in splitting the cluster MS. 
But in order to compare our models with the SYCLIST models, we use eq.\,11 in 
\citet{2013A&ARv..21...69R} to convert the critical velocity fraction of the SYCLIST models to the value under the MESA definition. 

It is worth noting that the SYCLIST stellar models undergo a relaxation phase at the very beginning of their MS evolution, lasting for a few percent of their MS lifetimes, during which their rotational velocities decrease dramatically as meridional circulation transports angular momentum from their outer layers to their inner layers \citep{Ekstrom2008}. The decrease of the surface rotational velocity is more substantial in models with higher initial velocities. In particular, the SYCLIST model formally labeled as $\omega_\mathrm{label}=\Omega_{\rm e}/\Omega_\mathrm{crit, SYCLIST}=0.9$ has $\Omega_{\rm e}/\Omega_\mathrm{crit,SYCLIST}\sim 0.8$ after such relaxation phase, which corresponds to $v_\mathrm{e}/v_\mathrm{crit, MESA} \sim 0.5$. Therefore, we stress that the SYCLIST models that are formally labeled as extreme rotation are in fact not rotating that fast.

\section{Input physics and assumptions in stellar models}\label{sec:method}
We use the single stellar models described in detail in \citet{2022NatAs...6..480W}, which are extended from the models in \citet{2019A&A...625A.132S}.
The models are computed by the detailed one-dimensional stellar evolution code MESA, version 12115, \citep{Paxton2011,Paxton2013,Paxton2015,Paxton2019}. The models include differential rotation, rotationally induced internal mixing and magnetic angular momentum transport.
In this section, we briefly summarize the most relevant assumptions in the following and compare them with those used in SYCLIST and MIST models.

The stellar structure of a rapidly rotating star deviates from spherical symmetry due to the presence of the centrifugal force. In a one-dimensional stellar evolution code, stellar structure equations are solved on the isobaric shells that are assumed to have constant angular velocities \citep{Meynet1997}. 
In the MESA code, two factors $f_T$ and $f_P$, are introduced to correct temperature and pressure of a rotating star, such that the regular form of the equations of nonrotating stars is retained \citep{Endal1976,Paxton2013}.  In the older versions of the MESA code, $f_T$ and $f_P$ are limited to specific values to ensure numerical stability, which limits the accuracy of computing the stellar models rotating at more than $60\%$ of their critical velocities.
However, in version 12115, which is the one we use, a new implementation of the centrifugal effects allows for a precise calculation of the stellar models rotating up to 90\% of their critical velocities \citep{Paxton2019}. 

We use the standard mixing-length theory to model convection with a mixing length parameter of $\alpha=l/H_{\mathrm{P}}=1.5$, where $H_\mathrm{P}$ is the local pressure scale height. The boundary of the convective core is determined by the Ledoux criterion. 
We adopt step overshooting that extends the convective zone by $\alpha_{\mathrm{OV}}H_{\mathrm{P}}$, where $\alpha_{\mathrm{OV}}$ varies with mass (see \citet{2021A&A...653A.144H,2022NatAs...6..480W} for detail).
In the SYCLIST modes, the same method is used with a fixed $\alpha_{\mathrm{OV}}$ of 0.1. In the MIST models, an exponentially decaying diffusion coefficient is applied, which is roughly equivalent to $\alpha_{\mathrm{OV}}=0.2$ in the context of step overshooting. A larger convective overshooting parameter will result in a larger convective core and thus a longer MS lifetime.
We adopt a semiconvection mixing efficiency parameter of $\alpha_{\mathrm{SC}}=10$ \citep{2019A&A...625A.132S}, and a thermohaline mixing efficiency of $\alpha_\mathrm{th}=1$ \citep{Cantiello2010}. 

Rotationally enhanced mixing is modeled as a diffusive process \citep{Heger2000}, with $f_c$ scaling the efficiency of composition mixing with respect to angular momentum transport, and $f_{\mu}$ describing 
the stabilizing effect of mean molecular weight gradients. We adopt values of $f_c=1/30$ \citep{Chaboyer1992,2016A&A...588A..50M} and $f_{\mu}=0.1$ \citep{2006A&A...460..199Y}, while the MIST models use the same $f_c$, but a different $f_{\mu}$ of 0.05.
A smaller $f_\mu$ value means a more efficient rotational mixing even in the presence of a stabilizing chemical composition gradient. 
The effects of the dynamical and secular shear instabilities, the Goldreich-Schubert-Fricke instability, and the Eddington-Sweet circulations are included. 
In the SYCLIST models, a more efficient diffusive-advective approach is taken when modeling rotational mixing, which 
makes their fast rotators hotter and mroe luminous than our models and than the MIST models with the same initial parameters. We implement the Tayler-Spruit dynamo \citep{Spruit2002} to transport angular momentum in our MESA models, which imposes a strong coupling between the contracting core and the expanding envelope during stars' MS evolution. As a consequence, our MESA models rotate nearly as solid bodies.
This process is not considered in the SYCLIST models, and consequently, the SYCLIST models rotate differentially and spin down more quickly at the stellar surface than the MESA models
\citep{Choi2016,2020A&A...633A.165H}. Differential rotation in the SYCLIST models induces shear mixing within the adjacent layers, which also leads to efficient rotational mixing.
Despite the difference of these parameters, they are all able to explain some observations but struggle with others \citep[see][for example]{Choi2016}, meaning that current observations cannot uniquely constrain the uncertain physics in stellar models.


We compute single-star models with both LMC-like metallicity $Z_\mathrm{LMC}=0.004841$ and SMC-like metallicity $Z_{\mathrm{SMC}}=0.002179$ according to \citet{Brott2011}. 
Our single-star models have initial masses between 1.05 and 19.95\mso, which securely covers the magnitude range of MS stars in the LMC cluster NGC\,1818, with intervals of $\log M/ \msun=0.04$. 
For each mass, we have nine evolutionary tracks with initial rotational velocities $W_{\mathrm i} = v_\mathrm{e}/v_\mathrm{crit, MESA}$ ranging from 0 to 0.8, in intervals of 0.1. Our models start from the zero-age MS (ZAMS) and end when their central helium is depleted.

We notice that our MESA models do not achieve thermal equilibrium initially, which causes wobbles in the stellar radius, and in turn affects their critical and rotational velocities. Unlike in the SYCLIST models, the change of the rotational velocity during this relaxation period is marginal in our MESA models, due to the implementation of a strong core-envelope coupling. 
To eliminate this uncertain phase, we redefine the ZAMS state as the time when 3\% of hydrogen is burnt in the core. We use the rotational velocity $v_{\mathrm e}/v_\mathrm {crit, MESA}$ at that time as the initial velocity $W_{\mathrm i}$. 
Then we perform interpolation to obtain stellar models with specified $W_{\mathrm i}$ values from 0 to 0.75, in intervals of 0.05.
To construct isochrones from our stellar models, we use the method described in \citet{2020ApJ...888L..12W} to obtain the stellar properties at certain ages.
We convert stellar luminosity and temperature to magnitude in the HST/WFC3 F814W filter and a color. For the color we use the difference between the magnitudes in the HST/WFC3 F336W and F814W filters, using the method described in \citet{2022NatAs...6..480W}.

To compare our models with the SYCLIST models, we first obtain the appropriate SYCLIST evolutionary tracks and isochrones, with marked $\omega_\mathrm{label}$ from 0.0 to 0.95, from their website. We then redefine the SYCLIST models' ZAMS phase and their initial rotational velocities using the method in Sec.\,\ref{sec:def}.
We also interpolate the SYCLIST models to obtain the models rotating at precisely the required values. 
For the MIST models, only the models with $W_{\rm i}=0.0$ and $W_{\rm i}=0.4$ are available to the public, thus we do not do any redefinition and interpolation to them. This does not affect our comparison, because the MIST models have the same definition of critical velocity as our MESA models, and meanwhile the effect of their initial relaxation phase before 3\% H consumption is negligible. We need to point out that in the SYCLIST models, metallicities $Z_{\rm LMC,\,SYCLIST}=0.006$ and $Z_{\rm SMC,\,SYCLIST}=0.002$ are tailored for LMC and SMC stars. The MIST models have multiple metallicities available. We chose two values that are closest to the adopted values in our MESA models, which are $Z_{\rm LMC,\,MIST}=0.0452$ and $Z_{\rm SMC,\,MIST}=0.00254$ for the LMC and SMC stars, respectively. We also need to stress that, apart from the different metallicities, the abundance ratio of the heavy elements are also different in the SYCLIST models and our MESA models. The SYCLIST models simply use solar-scaled abundance ratios, while our models follow the initial chemical compositions in \cite{Brott2011} that are intended to match the observations of the OB stars in the LMC and SMC in the VLT-FLAMES survey \citep{2005A&A...437..467E}.

\section{Results and discussions}\label{sec:Results}

\subsection{Rotationally induced color variation}\label{sec:Results_2}
We have mentioned in Sec.\,\ref{sec:intro} that the effects of rotation can change stellar effective temperature, luminosity, and lifetime. Previous studies usually inferred the spins for the red and blue MS stars by comparing the photometric observations with the isochrones of stellar models with different rotation \citep{Milone2018,2019ApJ...887..199G,2022NatAs...6..480W}. However, isochrone fitting is a nontrivial task due to many free parameters, including age, initial rotation, distance modulus, and reddening.
There is a more straightforward way to infer the spins of the red and blue MS stars in the first place, that is to calculate the color difference of the stellar models with different initial rotational velocities and then compare it with the observed color split of the red and blue MSs in young star clusters.  This experiment is not as sensitive to the adopted age, distance modulus, and reddening as isochrone fitting. After acquiring rotational velocities for the red and blue MS stars through this method, we inspect our results by performing isochrone fits to cluster stars in the CMD. 

\begin{figure*}[htbp]
\centering
\includegraphics[width=3.2 in]{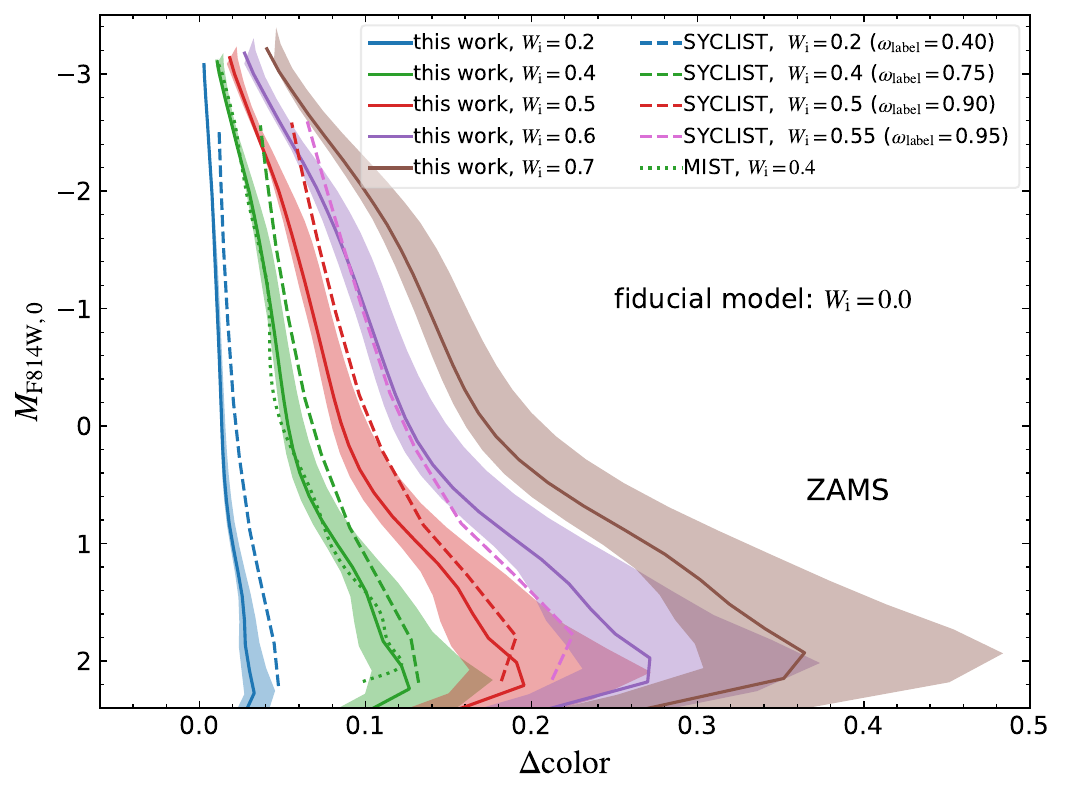}
\vspace{0.01cm}
\includegraphics[width=3.2 in]{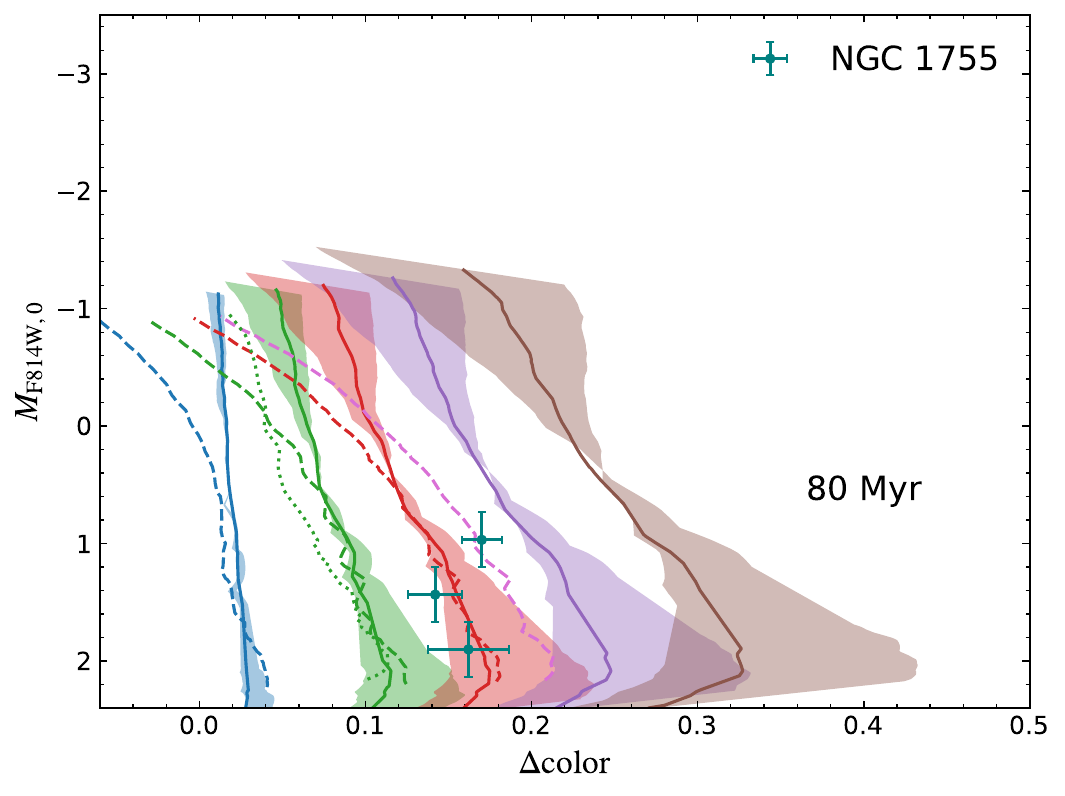}
\includegraphics[width=3.2 in]{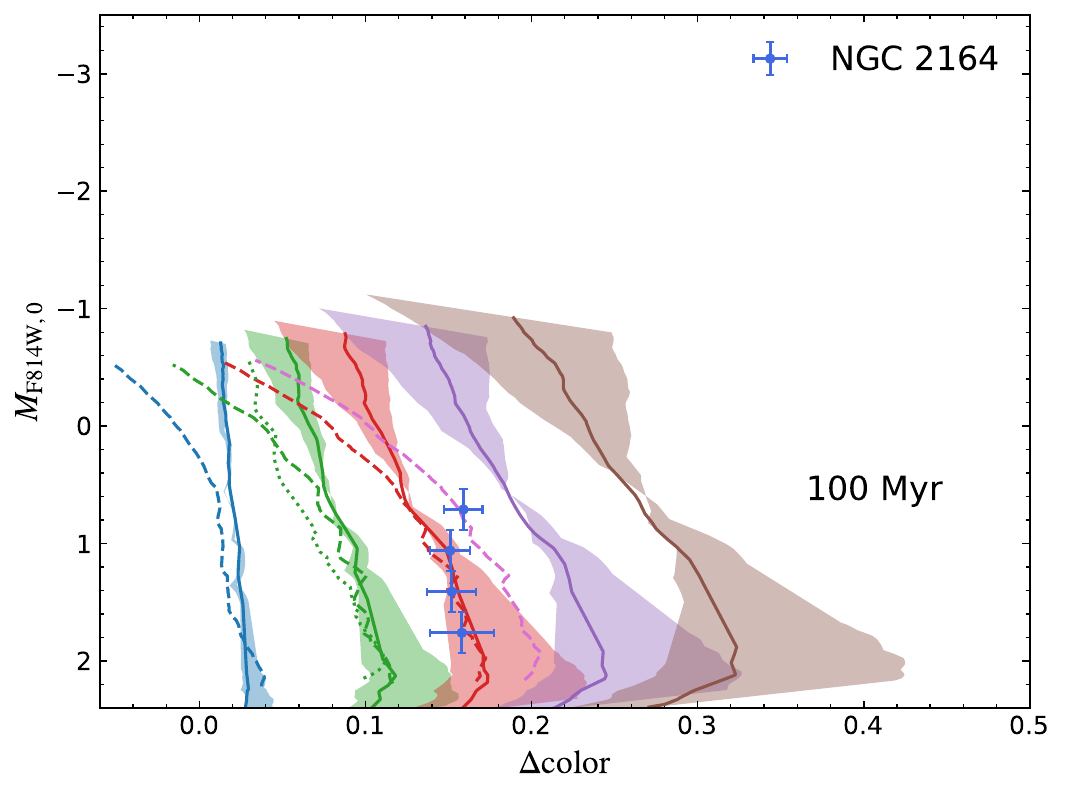}
\vspace{0.01cm}
\includegraphics[width=3.2 in]{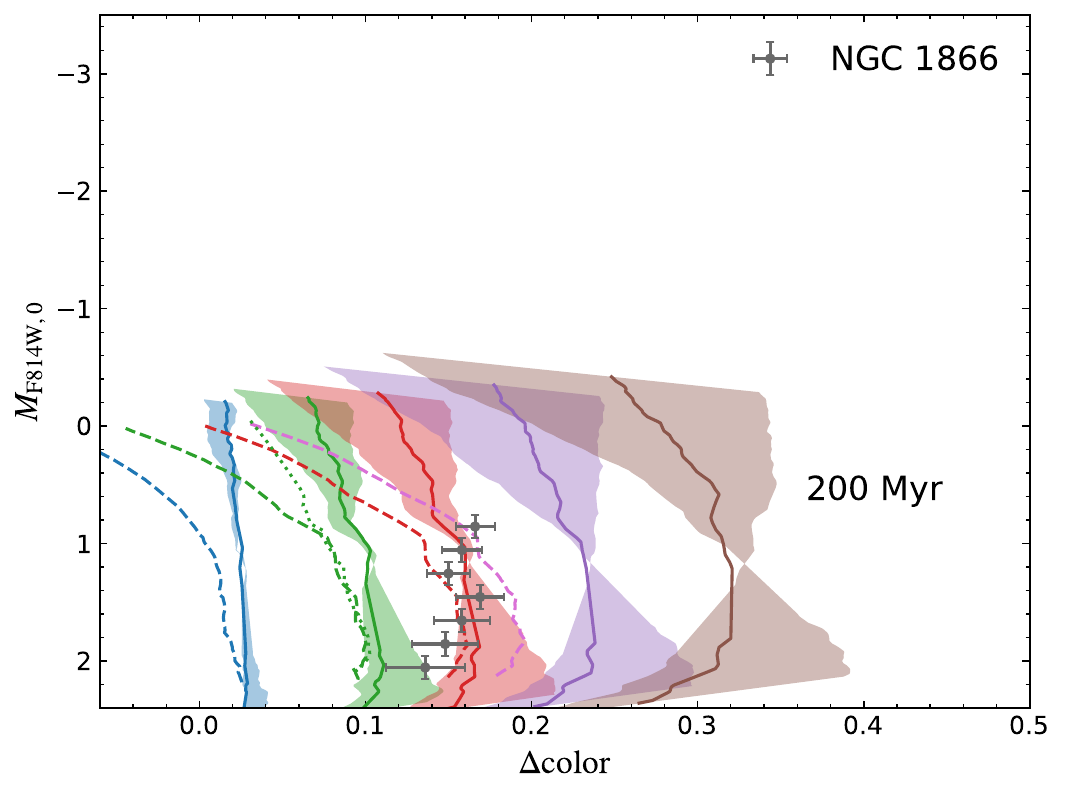}
\centering
\caption{Color difference between the rotating and nonrotating MS stellar models with an LMC metallicity as a function of absolute magnitude $M_\mathrm{F814W,0}$ at four different ages, which are the zero-age and the derived ages for three young LMC clusters NGC\,1755 (80\,Myr), NGC\,2164 (100\,Myr) and NGC\,1866 (200\,Myr) \citep{Milone2018}. 
The solid, dashed, and dotted lines denote our MESA models, the SYCLIST models and the MIST models, respectively, color-coded by their initial rotational rates. 
The formally labeled fractional critical angular velocities of the SYCLIST models in their web interface are shown in parentheses. Shaded areas show the $\Delta$color range occupied by stars with orientations that are between pole-on and equator-on.
The dots with error bars in the last three panels indicate the color difference between the red and blue MSs of the abovementioned three clusters (see fig.\,5 in \citealt{Milone2016} for NGC\,1755, fig.\,13 in \citealt{Milone2018} for NGC\,2164 and fig.\,6 in \citealt{Milone2017} for NGC\,1866, respectively). We only consider the unevolved stars below the cluster turnoff (see text). The error bars on the x-axis mean photometric errors while the error bars on the y-axis correspond to magnitude intervals. 
}
 \label{fig:color_split} 
\end{figure*}

To obtain the observed color split of the red and blue MSs in young star clusters, we utilize previous analyses on the fraction of stars in each MS in three clusters, namely NGC\,1755 \citep[see fig.\,5 in][]{Milone2016}, NGC\,2164 \citep[see fig.\,13 in][]{Milone2018} and NGC\,1866 \citep[see fig.\,6 in][]{Milone2017}. In these studies, the stellar density distribution as a function of color ($m_{\mathrm F336W}-m_{\mathrm F814W}$) has been scrutinized. We treat the color that has the highest density of the red and blue MS stars as the color of the corresponding population, and then calculate their color difference. We use the magnitude intervals given in these studies. We only take into account the unevolved and less-evolved observed stars, because the split MS feature is prominent among these stars. 
Meanwhile, focusing on the unevolved and less-evolved stars, which have not been affected by stellar evolution processes, allows us to identify the impact of stellar rotation. The results are shown by the dots with error bars in the last three panels of Fig.\,\ref{fig:color_split}, with different panels corresponding to different clusters. We convert the apparent magnitude of the observed stars to the absolute magnitude using the distance moduli and reddening values provided in \citet{Milone2018}. 

As to the stellar models, we calculate the color separation, $\Delta\,\mathrm{color}=\Delta(m_{\mathrm F336W}-m_{\mathrm F814W})$, of the isochrones of rotating single-star models and the isochrones of nonrotating single-star models in the color-magnitude diagram at four evolutionary times, that is zero age and the derived ages for the abovementioned three observed clusters. We include our MESA models, the SYCLIST models and the MIST models. The results are shown in Fig.\,\ref{fig:color_split}.

For our MESA models, we also examine the effect of the inclination of rotation axis on the color of the stellar models with eqs.\,44 and 45 in \cite{Paxton2019}. 
The results are shown by the shaded areas in Fig.\,\ref{fig:color_split}, with the left and right borderline denoting the pole-on and the equator-on case, respectively. 
As anticipated, at zero age, the projection effect of gravity darkening is more significant in more rapidly rotating stars. However, the projection effect of gravity darkening also contributes more at high magnitudes. We stress that it is merely a geometric consequence, since the isochrones of the stellar models with different initial rotational velocities have larger horizontal color differences at higher magnitudes (see the red shaded area in the right panel of Fig.\,\ref{fig:iso_fit}).
At older ages, the effect of gravity darkening becomes significant near the turn-off region. The reason is that gravity darkening either makes a star hotter and brighter (pole-on), or cooler and dimmer (equator-on); therefore, its net effect shifts a star in the CMD almost along the isochrone. Hence, the net effect of gravity darkening is only visible when an isochrone bends toward the right and becomes almost vertical in the turn-off region (also see the red shaded area in the right panel of Fig.\,\ref{fig:iso_fit}). 

Figure\,\ref{fig:color_split} yields three conclusions. Firstly, 
for the unevolved and less-evolved stellar models in which rotation only plays a role through the centrifugal force, the same initial velocity differences lead to nearly the same color variations in all the three model sets. Secondly, after the ZAMS, the color difference between the rotating and nonrotating models is different in different model sets. Among the three model sets, the SYCLIST models adopt the strongest rotational mixing efficiency (see Sec.\,\ref{sec:method} and Appendix\,\ref{sec:Appendix_A}). It can be seen that the $\Delta \mathrm{color}$ values of the SYCLIST models become smaller than that of our MESA models and the MIST models near the turnoff. In some cases, $\Delta \mathrm{color}$ is even negative, meaning that the corresponding SYCLIST rotating models are bluer than their nonrotating models. The adopted rotational mixing efficiency in the MIST models is slightly larger than that in our MESA models (see Sec.\,\ref{sec:method} and Appendix\,\ref{sec:Appendix_A}). As a consequence, the $\Delta \mathrm{color}$ values of the MIST models are between those of the SYCLIST models and our MESA models near the cluster turnoff.
The last conclusion is that the color difference between the models with $\sim$50\% of critical rotation, rather than the previously proposed near-critical rotation \citep{Antona2017,Milone2018}, and the models with zero rotation is adequate to retrieve the color split of the observed red and blue MSs in young star clusters. This is consistent with the previous results based on the MESA models \citep{2019ApJ...887..199G,2022NatAs...6..480W}. From the star formation point of view, half of the critical rotational velocity agrees with the results of hydrodynamic simulations that take into account the gravitational torques between stars and disks \citep{2011MNRAS.416..580L}.
We remind the reader that 50\% of the critical rotational velocity under the MESA definition roughly equals 60\% of critical linear velocity and 75\% of critical angular velocity under the definition in the SYCLIST models.

We notice that other combinations of initial rotational velocities are also able to explain the color difference of the observed red and blue MS stars. In Appendix\,\ref{sec:Appendix_B}, we show that $W_{\rm i}\sim0.55$ and $W_{\rm i}\sim0.6$ are required to explain the red MS stars, if the blue MS stars have $W_{\rm i}=0.2$ and $W_{\rm i}=0.3$, respectively. Meanwhile, \citet{2022NatAs...6..480W} showed that $W_{\rm i}\sim0.65$ and $W_{\rm i}\sim0.35$ can also explain the red and blue MSs in young star clusters.
We do not consider even faster rotation for the blue MS stars, otherwise it contradicts the spectroscopically measured low velocity of the blue MS stars in NGC\,1818 (see Sec.\,\ref{sec:v_n1818}).


\subsection{Isochrone fitting of the double main sequences in NGC\,1818}
\begin{figure*}[htbp]
\centering
\includegraphics[width=\linewidth]{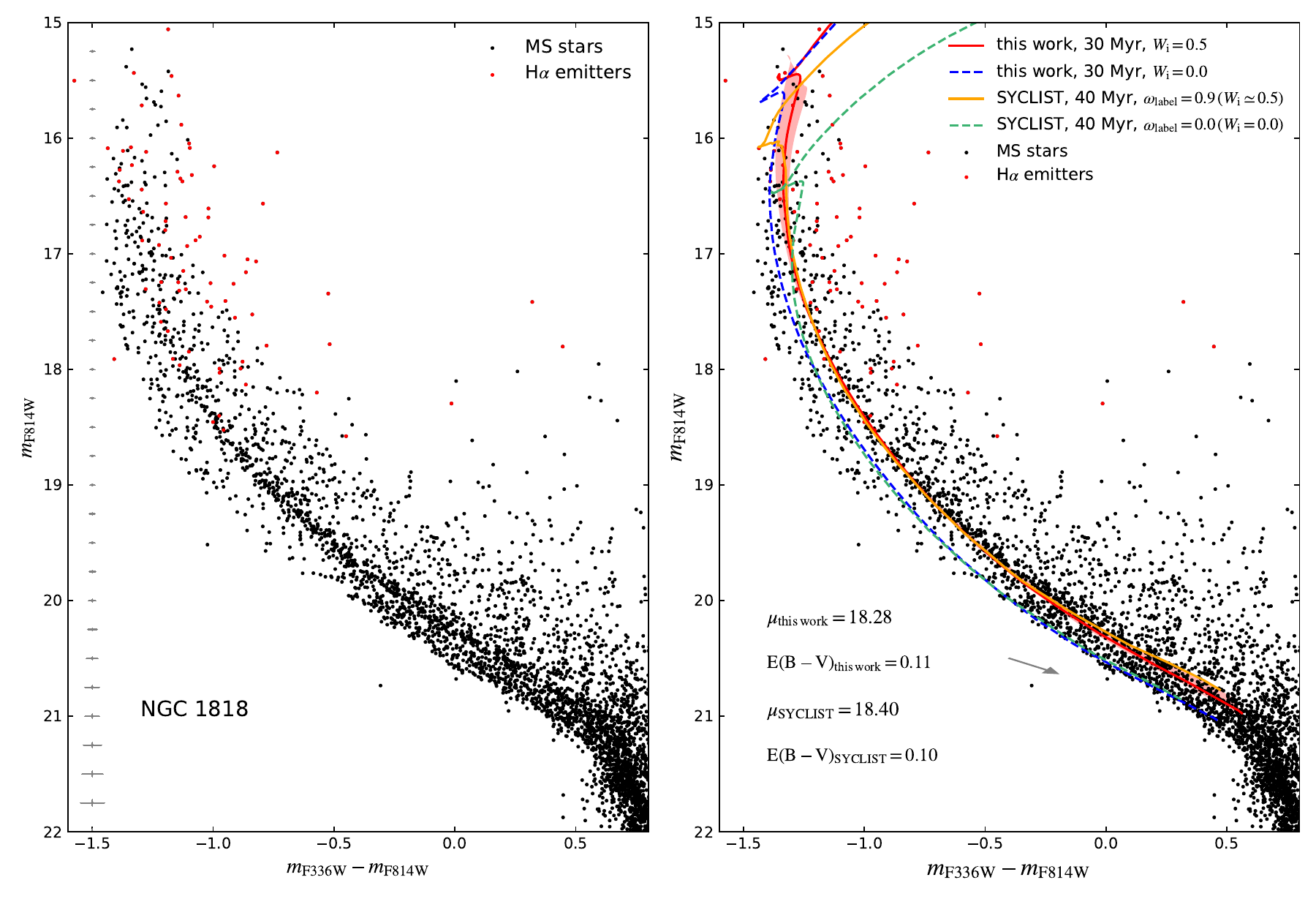}

\centering
\caption{Isochrone fitting to the main-sequence stars in the LMC cluster NGC,1818. Left: distribution of the main-sequence stars in NGC\,1818 observed by HST. The black and red dots are normal MS stars and the stars with a brightness excess in the H$\alpha$ narrow band filter, respectively. The gray error bars on the left indicate 1 $\sigma$ photometric uncertainties at corresponding magnitudes. Right: isochrone fitting to the red and blue MSs of NGC\,1818, using the SYCLIST models and our MESA models. For the SYCLIST models, we take the fit in \citet{Milone2018}, using 40\,Myr isochrones of the nonrotating models and the models with labeled 90\% of critical angular velocities (roughly equal to 50\% of critical linear velocities under the MESA definition) to fit the observed blue and red MSs, indicated by the solid orange and the dashed green lines, respectively. While for our MESA models, we employ 30\,Myr isochrones of the nonrotating models and the models with 50\% of critical linear velocities to fit the observed blue and red MSs, denoted by the solid red and the dashed blue lines, respectively. The adopted distance modulus and reddening in each fit are listed. The red shaded area depicts the projection effect (gravity darkening) on the red solid line. The small gray arrow shows how an isochrone would move if $E(B-V)=0.05$ is added. }
 \label{fig:iso_fit} 
\end{figure*}

In the following, we perform isochrone fitting to the red and blue MSs in the LMC cluster NGC\,1818 using our findings in Sec\,\ref{sec:Results_2} that these double MSs can be explained by the stellar models with $\sim$50\% and $\sim$0\% of their initial critical rotational velocities.
The left panel of Fig.\,\ref{fig:iso_fit} displays the distribution of the 
stars with (red dots) and without (black dots) H$\alpha$ excess in the MS region of NGC\,1818. 
We analyze this cluster because it exhibits a clear double MS feature, and has spectroscopic measurements for the rotational velocity of the stars in different MSs. 
In the right panel of this figure, we show isochrone fitting based on both our MESA models and the SYCLIST models.  
We first replicate the isochrone fitting shown in \citet{Milone2018}, in which 40\,Myr isochrones of the SYCLIST models with $\omega_\mathrm{label}=0.9$ and $\omega_\mathrm{label}=0.0$ are used. The adopted distance modulus and reddening are listed both in the figure and in Tab.\,\ref{tab1}. As we have mentioned, $\omega_\mathrm{label}=0.9$ in the SYCLIST models roughly correspond to $W_\mathrm{i}=0.5$ under the MESA definition. 

As to our MESA models, we adapt the isochrone age, distance modulus and reddening, such that the isochrone of our $W_\mathrm{i}=0.5$ MESA models (solid red line) fits the observed red MS as well as the SYCLIST isochrone, in a visual inspection. We find that a younger age is required when using our MESA models to achieve a satisfactory fit compared to using the SYCLIST models. This is because extensive rotational mixing makes fast-rotating SYCLIST models appear younger, indicating again that different physics assumptions in stellar models can give rise to different results in cluster age estimation. In this figure, we also display the impact of gravity darkening and inclination on our MESA isochrone of the fast-rotating-star models in the CMD. As explained in Sec.\,\ref{sec:Results_2}, the influence of gravity darkening in the CMD is only visible in the cluster turn-off area.

We notice that the adopted distance modulus $\mu=18.28$ for our MESA models is smaller than the measured value for the LMC ($\mu=18.49\pm 0.06$) \citep{2013Natur.495...76P,2016ApJ...832..176I}. Distance moduli of 18.49 and 18.28 correspond to distances of 49.9\,kpc and 45.3\,kpc. The diameter of the LMC estimated in \citet{2018A&A...616A..12G} is around 5.2\,kpc, meaning that the spatial extent of the LMC clusters is inconsistent with the small value of the adopted distance modulus. We stress here that the adopted distance modulus is sensitive to metallicity and chemical composition of the employed stellar models. Meanwhile, the adopted bolometric correction table also affects the derived distance modulus. Moreover, compared to the distance modulus, the position of an isochrone in the CMD depends more on the adopted reddening. 
The mean measured reddening of the LMC is $E(B-V)=0.113\pm0.060$ mag \citep{2019A&A...628A..51J}.
The gray arrow in the right panel of Fig.\,\ref{fig:iso_fit}
demonstrates in which direction the isochrones will move in the CMD if reddening is increased by 0.05. In summary, we state that our adopted distance modulus and reddening are in the acceptable range.

Figure\,\ref{fig:iso_fit} supports our previous result that, in both our MESA models and the SYCLIST models, 50\% and 0\% of critical rotation can explain the best discernible part of the observed red and blue MSs below the cluster turnoff. Our MESA models and the SYCLIST models show different behavior in the turn-off area.
The isochrone of the fast-rotating SYCLIST models crosses with that of the slowly-rotating stellar models, as a consequence of strong rotational mixing. In contrast, the isochrone of our fast-rotating MESA models is always redder than the isochrone of our slowly-rotating-star models. It seems that our MESA isochrones match the spread distribution of the turn-off stars better, as the SYCLIST isochrones miss the observed blue MS stars between 16 and 17.5th magnitude in the F814W filter.
However, considering the fact that the cluster turn-off area may contain a mixture of stars, including single stars, pre- and post-interaction binaries,  we cannot assess these two model sets based on current photometric observations. 
A plausible way to constrain rotational mixing is to measure the chemical abundance of the H-burning products at the stellar surface. We explore the surface He and N enrichment of our MESA models and the SYCLIST models in Sec.\,\ref{sec:surf_abundance}. 

We show in Appendix\,\ref{sec:Appendix_B} that other combinations of rotational velocities can produce equally satisfying isochrone fittings as Fig.\,\ref{fig:iso_fit}.
This again means that the precise rotational velocity of the cluster stars cannot be solely determined by photometric observations. 
Nevertheless, the conclusion that moderate and slow rotation is required to explain the split MSs in young star clusters is rigid. 

The stars redder than the isochrones of the fast-rotating-star models can be understood by the unresolved binaries containing two fast-rotating components. The presence of a secondary star shifts an observed star to a redder and brighter position in the CMD. The H$\alpha$ emitters, which are widely considered to rotate at or close to their critical rotational velocities, occupy the reddest part of the CMD, extending $\sim 2$ magnitudes from the cluster turnoff. We discuss the formation of these stars in Sec.\,\ref{sec:be stars}. At last, the stars bluer than the isochrones of the slowly-rotating-star models in Fig.\,\ref{fig:iso_fit} are explained as MS merger products \citep{2020ApJ...888L..12W,2022NatAs...6..480W}. In addition to their slow rotation, MS merger products appear bluer also because of rejuvenation.

\begin{table*}
\begin{center}
\caption{Models and parameters used in isochrone fittings to NGC\,1818.}
\begin{tabular}{ l c c c c c c c} 
 \toprule
Fitting model        & $W_{\rm i} $ for the red MS &  $W_{\rm i}$ for the blue MS &   Age (Myr)  & $\mu$  &    $E(B-V)$ \\
 \midrule
Our MESA models           & 0.50   &0.00  &   30         & 18.28 &  0.11\\
Our MESA models           & 0.55   &0.20   &   35         & 18.29 &  0.09\\
Our MESA models           & 0.60   &0.30   &   35         & 18.31 &  0.09\\
Our MESA models         & 0.65     &0.35  &  40        &18.31    &0.07 \\
SYCLIST models          & 0.50   &0.00     & 40        & 18.40 &  0.10 \\
 \bottomrule
  \label{tab1} 
\end{tabular}
\end{center}
\end{table*}



\subsection{Surface chemical composition}\label{sec:surf_abundance}
Massive stars burn H through the CNO cycle, in which the rate of the reaction between $^{14}$N and $^1$H is the slowest. This leads to nitrogen enhancement in the layers where nuclear burning takes (or has taken) place. Rotational mixing may take these layers that are enriched in $^{14}$N, as well as the H burning ashes He, to the stellar surface. Hence, surface chemical composition abundance can be used as a probe of the strength of rotational mixing.
In Fig.\,\ref{fig:iso_fit}, we show that our MESA isochrones and the SYCLIST isochrones can fit the split MS in young star clusters equally well, regardless of the different adopted rotational mixing efficiencies. In this section we seek clues to assess these two model sets --- or, being more precise, to constrain the strength of rotational mixing --- by studying the surface abundance of nitrogen and helium in different stellar models. 

\begin{figure*} 
	\includegraphics[width=0.98\linewidth]{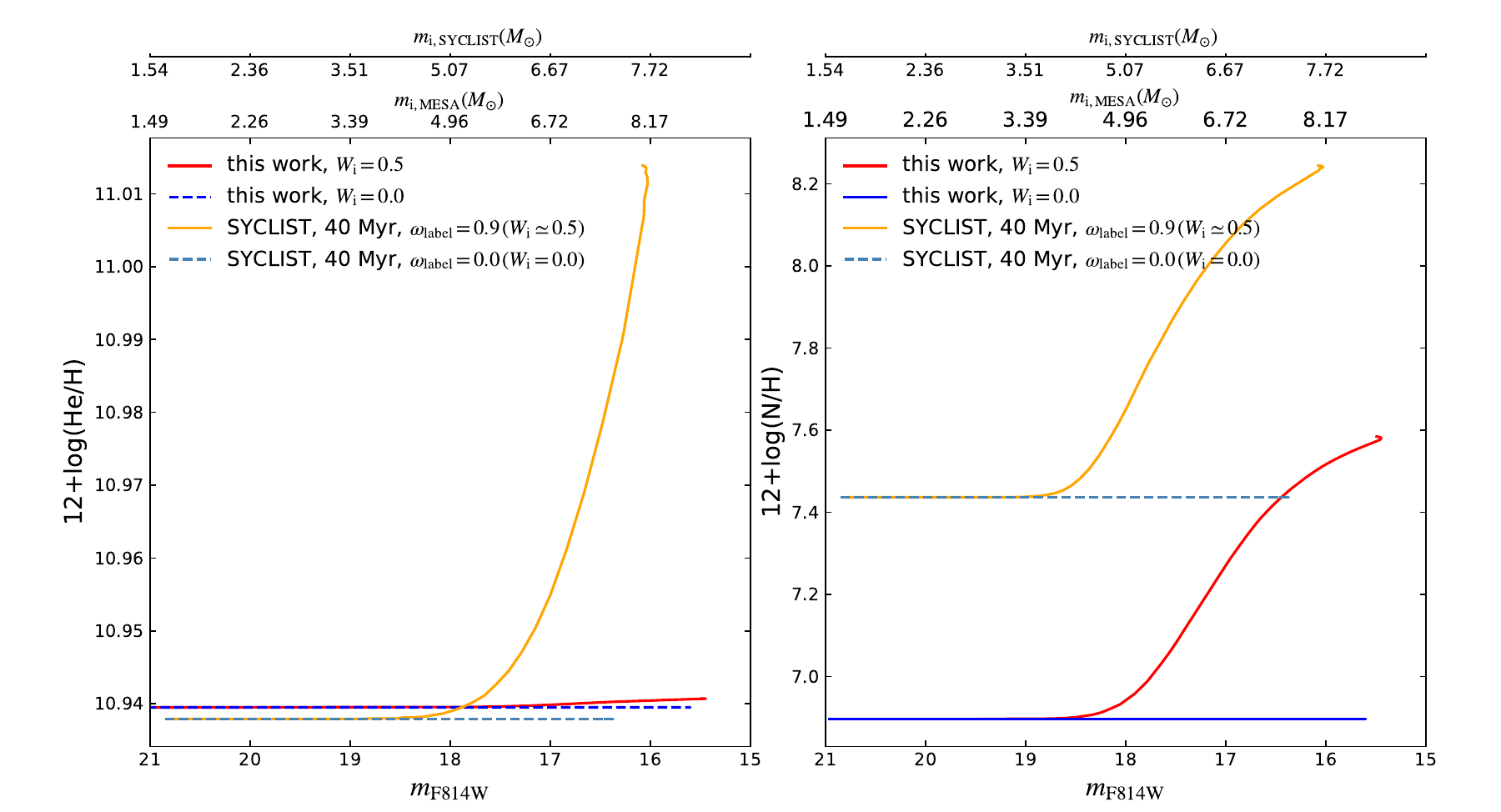}
	\centering
	\caption{Current surface He (left) and N (right) abundances as a function of magnitude for our MESA models and the SYCLIST models that are used to build the isochrones in Fig.\,\ref{fig:iso_fit}. The upper two x-axes show the corresponding mass derived from our MESA models and the SYCLIST models with rapid rotation.}
	\label{fig:app_c1} 
\end{figure*}

Figure\,\ref{fig:app_c1} shows the surface He and N abundances of the stellar models that are employed in building the isochrones in the right panel of Fig.\,\ref{fig:iso_fit}.
It can be seen that the rotationally induced surface N enhancement of our MESA models and the SYCLIST models is almost the same ($\sim 10\%$) at the cluster turnoff.
However, a significant enhancement of surface He abundance is only seen in the SYCLIST fast-rotating models. The different behavior of the surface He and N enrichment may be due to the fact that He is produced more slowly than N. 
When He is produced, a strong chemical composition gradient is created on top of the convective core, which suppresses further mixing of He to the stellar surface in our MESA models, but plays a less important role in the SYCLIST models. The significant surface He enhancement in the SYCLIST fast-rotating models near the cluster turnoff explains their higher temperature. 
Despite a scarcity of detailed measurements of He abundance of the stars in young star clusters, \cite{2020AJ....159..152C} found similar surface He abundances for the red and blue MS stars in the turn-off area of the young SMC star cluster NGC\,330, which may argue against efficient rotational mixing. Nevertheless, their results should be taken with a grain of salt, given their low S/N, resolution and small sample size. Further observations on more stars in more star clusters are in high demand for our understanding of rotational mixing. We note that recent asteroseismology studies inferred near-rigid rotation in late-B and AF stars \citep{2021RvMP...93a5001A}, and a small amount of differential rotation in massive stars \citep{2020FrASS...7...70B}, which seems to support our MESA models with efficient angular momentum transfer in the stellar interior.

\begin{figure*} 
	\includegraphics[width=0.5\linewidth]{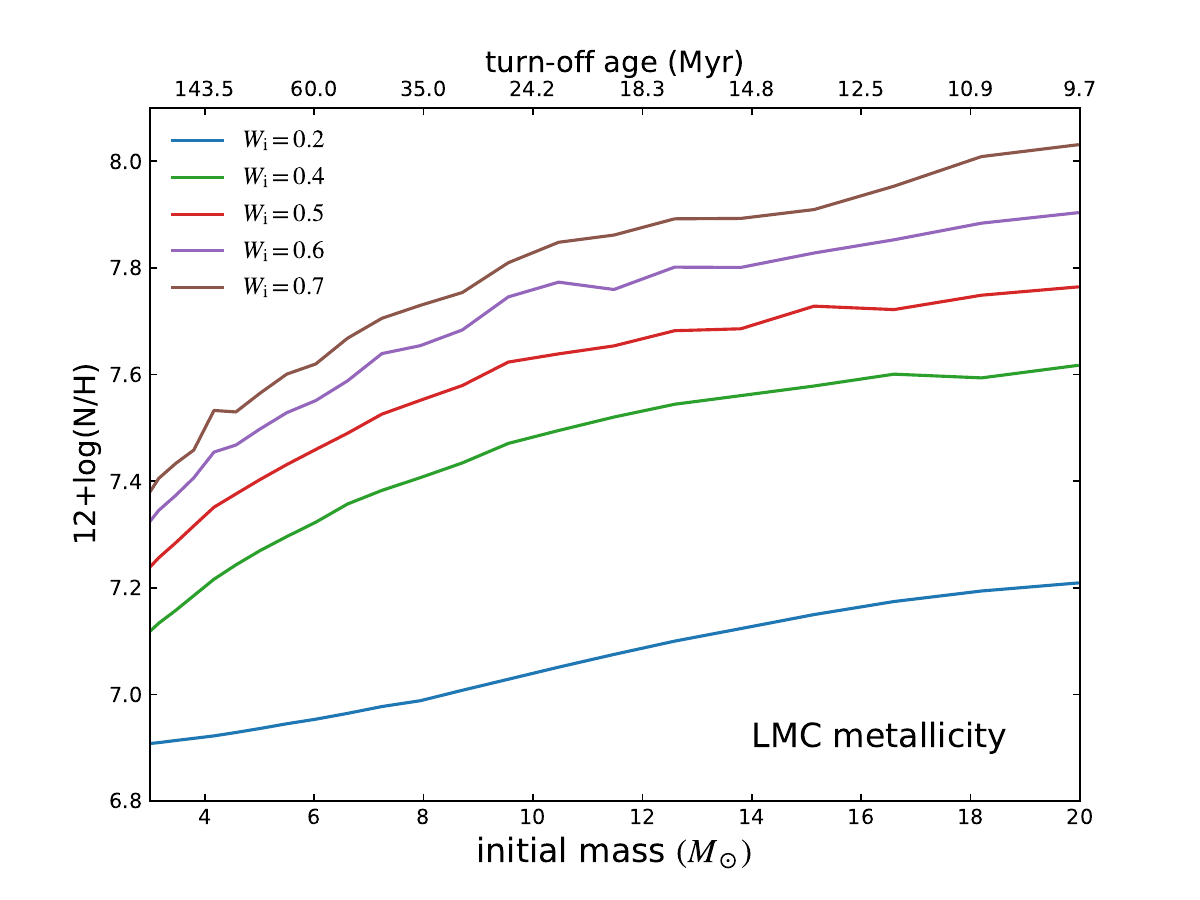}
	\includegraphics[width=0.5\linewidth]{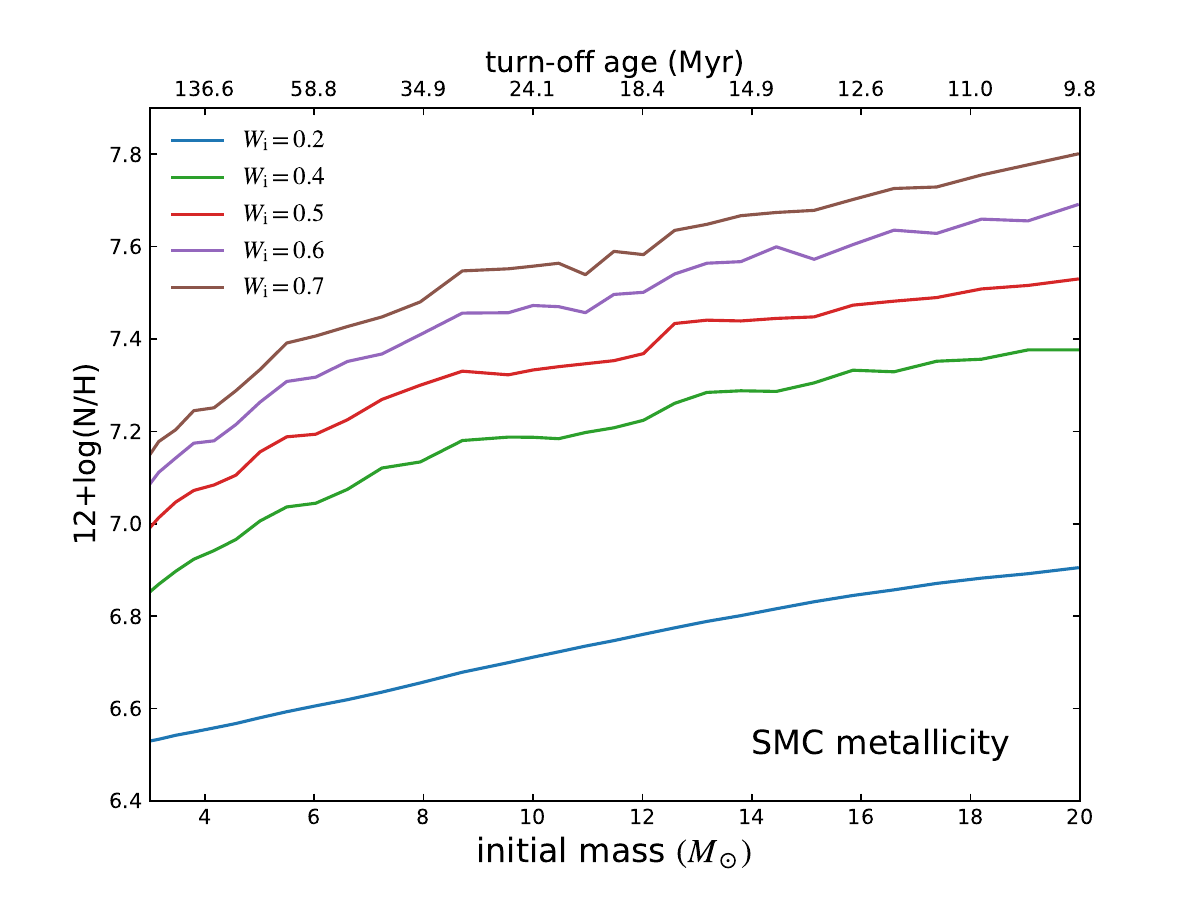}

	\centering
	\caption{Surface N abundances for our MESA models with the LMC metallicity (left) and the SMC metallicity (right) at the end of their main sequences when 1\% of H is left in the stellar center. The x-axis shows the initial mass of the stellar models, with the corresponding cluster turn-off ages shown on the top x-axis. Solid lines with different colors indicate different initial rotational rates.}
	\label{fig:app_c2} 
\end{figure*}

We have shown that stellar models with $W_\mathrm{i}=0.5$ exhibit a remarkable surface N enrichment at the end of their MS evolution, in both model sets. In the following, we examine whether the stellar models with other initial velocities also have a surface N enrichment at their TAMS. The results are shown in Fig.\,\ref{fig:app_c2}, with the left and the right panel corresponding to our MESA TAMS models appropriate for the LMC and SMC metallicities, respectively. 
It can be seen that the stellar models with higher initial velocities and larger initial masses will have larger surface N abundances at their TAMS, which agrees well with the findings in \citet{2020A&A...633A.165H}. Our results indicate that single stars born with spins larger than 40\% of their critical values will exhibit a significant N enrichment at the end of their MS evolution.

\subsection{Comparison with the spectroscopic measurements of the rotational velocity of the stars in NGC\,1818}\label{sec:v_n1818}

\begin{figure*} 
	\includegraphics[width=0.98\linewidth]{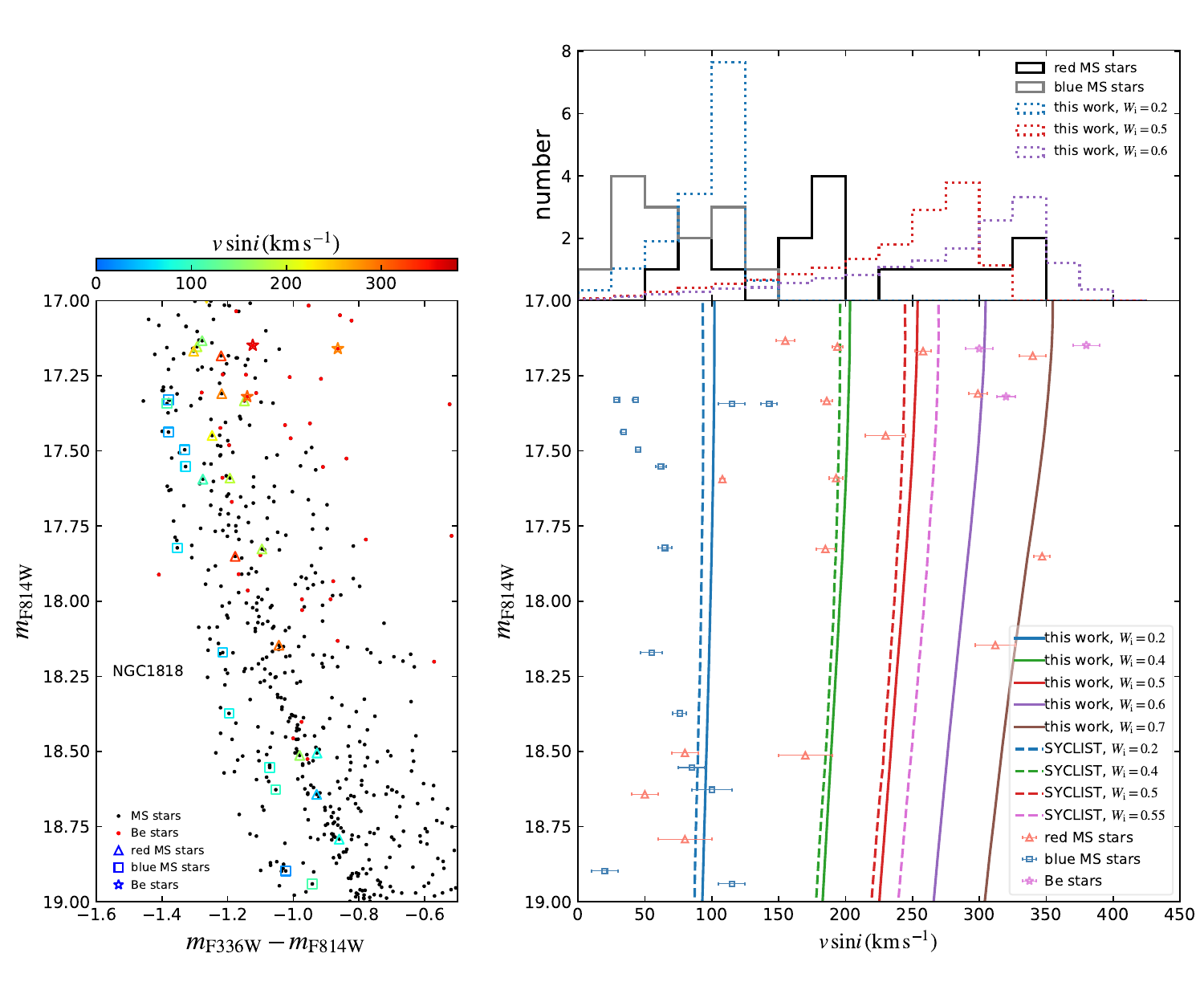}
	\centering
	\caption{Comparison between the predicted and the observed rotational velocities. Left: a zoom-in image of the main-sequence stars in NGC\,1818 which have magnitudes between 17 and 19. The black and red dots correspond to the normal main-sequence stars and H$\alpha$ emitters, respectively. The open symbols mark the stars that have spectroscopic rotational velocity measurements \citep{Marino2018}, with color indicating their measured projected rotational velocity $v{\rm sin}\,i$ values. Here $i$ is the inclination. The red main-sequence stars, blue main-sequence stars, and Be stars with H$\alpha$ emission lines are designated by triangles, squares and asterisks, respectively. Lower right: comparison between the model predicted average surface equatorial velocity and the spectroscopic measurements. The open symbols with error bars correspond to the observations, using the same symbol types as the left panel. The solid lines and dashed lines represent the average surface equatorial velocity (times $\pi/4$ to account for a random orientation) of our MESA models at 30\,Myr and the SYCLIST models at 40\,Myr (the ages used in our isochrone fitting). Color coding is the same as in Fig.\,\ref{fig:color_split}.
Upper right: comparison between our theoretical predicted $v{\rm sin}i$ distributions (color dotted steps) and the observed $v{\rm sin}i$ distributions of the red (solid black steps) and the blue (solid gray steps) main-sequence stars. For the theoretical predictions, we perform a Monte Carlo simulation by assuming a random orientation angle distribution and normalize the results to a total number of 15 for each initial rotation, which is similar to the number of the blue (14) and red main-sequence (16) stars that have spectroscopic observations.}
	\label{fig:1818_vrot} 
\end{figure*}

By studying photometric observations, we have drawn the conclusion that the red and blue MSs in young star clusters are comprised of the stars with natal spins of 50\%-65\% and 0\%-35\% of the critical velocities, respectively, by studying photometric observations.
Recently, several high-resolution spectroscopic studies have been performed in young and intermediate-age clusters, providing an unprecedented opportunity to directly measure the rotational rate of the red and the blue MS stars \citep{Dupree2017,2018MNRAS.480.1689K,Bastian2018b,Marino2018,2020MNRAS.492.2177K,2021A&A...652A..70B}. Among them, NGC\,1818, the cluster we study in this work, is so far, the youngest cluster that has detailed velocity measurements for the stars in different MSs \citep{Marino2018}.

The left panel of Fig.\,\ref{fig:1818_vrot} is a zoomed-in version of the CMD of the NGC\,1818 MS stars which have magnitudes between 17 and 19. We mark the MS stars whose velocities have been measured spectroscopically in \cite{Marino2018}. 
In the lower right panel, we display the observed $v\,{\rm{sin}}i$ of these MS stars as a function of magnitude, and compare them with the stellar models. 
The average measured velocities for the red and the blue MS stars are $202\pm23\,\mathrm{km\,s^{-1}}$ and $71\pm10\,\mathrm{km\,s^{-1}}$, respectively, with the red MS stars having a more dispersed velocity distribution. This provides compelling evidence that rotation is responsible for the split MS, with the red MS stars rotating faster than the blue MS stars. The large velocity dispersion of the red MS stars may be attributed to their large intrinsic velocities in combination with the effect of inclination. 
It is worth noting that all the three red MS stars fainter than $\sim$18.5th magnitudes are reported to be slow rotators. \cite{Marino2018} argued that the accuracy of the velocity measurement for these faint stars may be affected by their large systematic errors. The Be stars are shown to have the largest $v\,{\rm{sin}}i$ values on average.
In the upper right panel of this figure, the solid steps show the sum of the number of the observed MS stars in different velocity bins in each population.

To compare with the observations, we display the velocities of our MESA models and the SYCLIST models in the right panels of Fig.\,\ref{fig:1818_vrot}. In the lower right panel, the red lines demonstrate the average surface rotational velocity (times $\pi/4$ to account for a random orientation) of the $W\mathrm{i}=0.5$ stellar models along the isochrones that are used to fit the red MS in the right panel of Fig.\,\ref{fig:iso_fit}, as a function of magnitude. 
Then we also display the results for the stellar models that have the same ages but different initial velocities (see other color lines). Consistent with our findings from photometric observations, the current surface velocity of the stellar models with $W_{\rm i}\sim0.5$ agrees well with the mean observed rotational velocity of the red MS stars ($202\pm23\,\mathrm{km\,s^{-1}}$). But the three fastest rotating red MS stars can only be explained by the stellar models with initial velocities larger than $\sim 70\%$ of their critical values, assuming $\sin i=\pi/4$. As to the blue MS stars, the spectroscopic observations report slow, yet clearly non-zero velocities ($\sim 10-35\%$ of their critical velocities).
In the upper right panel of Fig.\,\ref{fig:1818_vrot}, we show the number distribution of our MESA models as a function of their velocities with dashed steps. To do this, we again assume a random orientation for our models. We employ our MESA models with $W_{\rm i}=0.2$, $0.5$ and $0.6$ and consider a Salpeter IMF with the exponent of -2.35 \citep{Salpeter1955}. We generate $10^6$ stellar models and then normalize them to a total number of 15 for each rotational velocity. This number is chosen such that it is similar to the number of the spectroscopically studied red (16) and blue (14) MS stars. 
Comparing with the observations, we conclude that current velocity of our stellar models with $W_{\rm i} \sim 0.5-0.6$ and the stellar models with $W_{\rm i} \sim 0.2$ cover the velocity range of the observed red and blue MS stars well, respectively. 
The mismatch between the peak of the star model distribution and the observation may be attributed to the small sample size of this observation. Hence, further spectroscopic observations of more stars in this cluster and the stars in other clusters are in high demand to thoroughly constrain the rotation of the cluster MS stars. 


\subsection{Implications for the origin of Be stars}\label{sec:be stars}
\begin{figure*}[htbp]
\centering
\includegraphics[width=\linewidth]{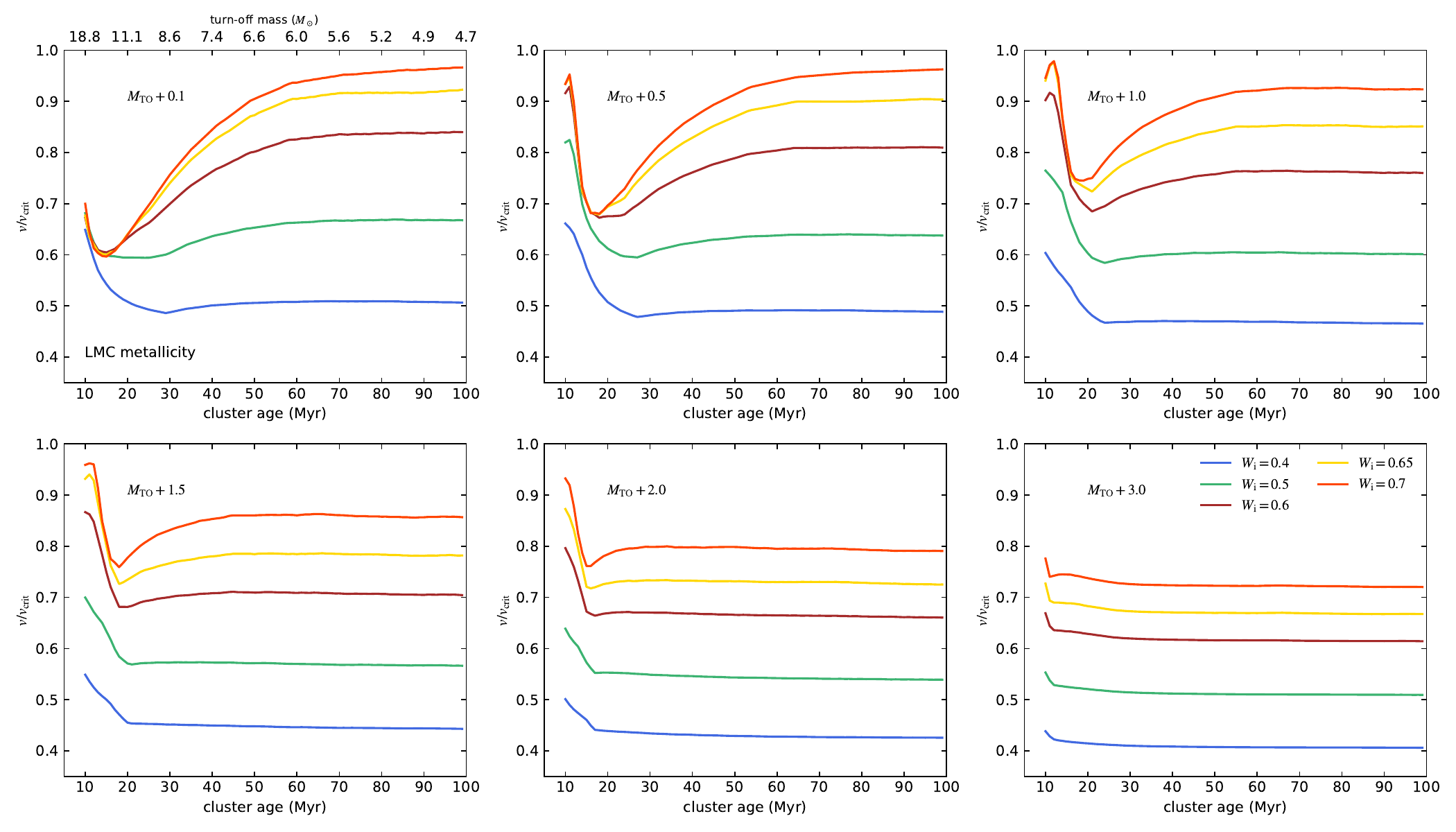}
\centering
\caption{Surface critical velocity fraction $v/v_\mathrm{crit}$ of our LMC models with different magnitudes as a function of cluster age. Each panel shows the result for the stellar models with a given magnitude, indicated in the legend. $M_\mathrm{TO}$ denotes the absolute $F814W$ magnitude of the cluster turnoff. In each panel, different colors correspond to the stellar models with different initial rotational velocities. We show the turn-off mass of the clusters with corresponding ages on the top x-axis in the first panel. 
}
 \label{fig:v_vcrit_age} 
\end{figure*}

Currently, two plausible ways have been proposed for the formation of Be stars. The first one is the so-called single-star channel, in which a rotating single star achieves near-critical rotation toward the end of its MS evolution, due to angular momentum transport from the contracting core to the expanding envelope \citep{Ekstrom2008,2020A&A...633A.165H}. The second way is the so-called binary channel, in which the accretor in a binary system spins up during Roche-lobe overflow \citep{1991A&A...241..419P,Langer2012,2020ApJ...888L..12W}. The detection of Be/X-ray binaries comprised of a Be star and a compact object \citep{2005A&AT...24..151R,2006A&A...455.1165L} and Be+sdO binaries \citep{2008ApJ...686.1280P,2017ApJ...843...60W,2018A&A...615A..30S,2021AJ....161..248W} provide direct evidence for the binary channel. However, which fraction of the Be stars has a binary origin is still unknown \citep{1991A&A...241..419P,1997A&A...322..116V,2014ApJ...796...37S}.
Meanwhile, \citet{2020A&A...641A..42B} found a lack of MS companions to Be stars, which indicates that the binary channel could dominate Be star formation. 
The H$\alpha$ emitters detected in young star clusters may  
deliver new insights into the origin of Be stars. 
\cite{2020A&A...633A.165H} stated that the detection of Be stars several magnitudes fainter than the cluster turnoff may advocate the major role of binary interaction in producing Be stars, because the single-star channel can only lead to the Be stars near the cluster turnoff, unless stars are born with extremely high spins. Because they adopted an initial rotational velocity distribution based on the observations of early B stars in the LMC \citep{Dufton2013} and they used different stellar models \citep{Brott2011} from ours, we find it helpful to reinvestigate the contribution of single-star evolution in producing Be stars in young star clusters with our MESA models, considering our findings that the red MS stars are born with 50--65\% of their critical rotational velocities. 
In Fig.\,\ref{fig:v_vcrit_age} we show the surface critical rotational velocity fraction $v/v_\mathrm{crit}$ of our LMC models whose magnitudes are between 0.1 and 3 mags (in the F814W filter) below the cluster turnoff. We only consider the stellar models which have initial spins higher than 40\% of their critical values, because the stellar models rotating more slowly only change their rotational velocities marginally during their MS evolution and cannot give rise to Be stars. We consider star clusters with ages from 10 to 100\,Myr, which are covered well by our stellar models.
Nevertheless, our conclusion should apply to star clusters as old as those with cluster ages corresponding to a turn-off mass of around 2.5\,$M_\odot$, below which magnetic braking may play a vital role in determining stars' initial rotational velocities.

The last panel of Figure\,\ref{fig:v_vcrit_age} demonstrates that the stellar models three magnitudes below the cluster turn-off luminosity almost retain their initial rotational velocities. Only the stellar models in clusters younger than $\sim$30\,Myr evolve to slightly higher critical rotational velocity fractions, because the stellar models three magnitudes below the cluster turnoff in such young star clusters are in later evolutionary phases, that is, have consumed more hydrogen, compared to those in older star clusters. 
The other panels in this figure represent the results for brighter stars. It can be seen that in clusters older than $\sim$60\,Myr, $v/v_\mathrm{crit}$ increases with stellar brightness monotonically and even reach close to 1 near the cluster turnoff. While in clusters between 10 and 60\,Myr, there is a ``U-shape'' feature for the stellar models with $W_\mathrm{i}\geq0.6$. The right side of the ``U-shape'' feature is caused by strong stellar winds due to the bi-stability jump, which have a larger impact on more massive stars. While the left side of the ``U-shape'' feature is a combined effect of a later evolutionary phase of the corresponding stellar models and a smaller impact of the enhanced stellar winds due to the bi-stability jump, because the related massive stellar models are only affected by the bi-stability jump at the very end of their MS evolution. However, whether the bi-stability jump indeed exists and how it affects stellar evolution is still debated \citep{2022arXiv220308218B}. We explain this in more detail in Appendix\,\ref{sec:Appendix_D}.

Figure\,\ref{fig:v_vcrit_age} shows that the contribution of single-star evolution to the Be stars in young star clusters depends on how fast single stars initially rotate and how fast B stars have to rotate to become Be stars. 
Unlike the binary channel that mainly produces near critically rotating Be stars \citep{2020ApJ...888L..12W}, the single-star channel gives rise to stars of various rotational velocities. Given the fact that how fast Be stars rotate is disputed \citep{2001A&A...368..912Y,2005ApJ...634..585C,2016A&A...595A.132Z}, we only provide some instructive discussions.
We have concluded that the red MS stars in young star clusters have 50--65\% of their critical rotational velocities. Figure\,\ref{fig:v_vcrit_age} illustrates that the stellar models with $W_\mathrm{i}=0.5$ cannot give rise to Be stars if we consider 70\% of the critical rotational velocity as the threshold for a star to be a Be star, unless in very young star clusters ($\sim$10\,Myr). We do not consider an even lower velocity threshold for Be stars, because otherwise the red MS stars may be born as Be stars according to their derived initial spins, which contradicts the observations that H$\alpha$ emitters are only found within 2-3 magnitudes below the cluster turnoff. If the initial rotational velocity for the red MS stars can be as high as 65\% of their critical values, single star evolution is able to explain the Be stars as faint as 2 magnitudes below the cluster turnoff, given a Be star velocity threshold of 70\% of critical rotational velocity. But it becomes hard to explain the Be stars with the single-star channel if the Be star velocity threshold is higher than 80\% of critical rotation. Near the cluster turn-off region it is stellar winds that prevent stellar models from being Be stars, while below the cluster turn-off stars do not have enough time to evolve to very high spins. 

In summary we find that, from the theoretical side, the significance of the role of single-star evolution in producing Be stars in young star clusters depends on the uncertain stellar wind and the uncertain rotational velocity required to form Be stars. Though 
\citet{2020ApJ...888L..12W,2022NatAs...6..480W} showed that binary evolution can explain the luminosity range of the observed H$\alpha$ emitters in young star clusters, and \citet{2021A&A...653A.144H} found that binary interaction alone is able to explain the luminosity range and number fraction of the observed Be stars in young star clusters based on a simple analytical analysis, we cannot exclude the possibility that the single-star channel produces Be stars. Accurate estimation of the spins of Be stars should be useful since, as we have mentioned above, Be stars stemming from the single-star channel should have a larger spin range than those originating from the binary channel. In addition, Be stars originating from the single-star channel usually have a visible surface N enrichment (see Sec.\,\ref{sec:surf_abundance}). On the contrary, the Be stars formed in binary systems may bear a normal N abundance, because the progenitors of these Be stars, that is the accretors, rotate at normal speed until mass transfer starts, at which time a strong chemical composition gradient that can impede a further mixing has been established. But whether binary-evolution-induced Be stars have N enrichment may also depend on the accretion efficiency, as the N enriched layers of the donor stars may remain on the accretors.

We show the results for the SMC models in Fig.\,\ref{fig:app_d1}. We attain a similar conclusion in the SMC models, but, in general, the SMC models reach higher velocities than their LMC counterparts as a consequence of weaker wind mass loss. 

\section{Summary \& conclusions}\label{sec:conclusion}
There is a growing consensus that a bimodal distribution of rotation rates causes the split MSs in young star clusters. Previous studies proposed that the red and blue MSs are comprised of near-critical rotating stars and nonrotating stars, respectively, by comparing the SYCLIST models with photometric observations.
However, the suggestion that most of the stars are born with near-critical rotation contradicts the spectroscopically measured velocity of both cluster and field stars. \citet{2019ApJ...887..199G} and \citet{2022NatAs...6..480W} argued that the red MS stars should rotate at around half of their critical velocities. In this work, we reinvestigated how much rotation is needed to explain the color split of the observed double MSs in young star clusters, considering our newly computed MESA models, the SYCLIST and the MIST models. We pointed out that it is a misconception that the SYCLIST models formally labeled as rotating at 90\% of their critical angular velocities are rotating close to their critical rotational velocity, because of the early relaxation stage of the models and different definitions of critical rotation. We indeed found consistency between our MESA models and the SYCLIST models. Moreover, we found that the cluster MS stars follow a bi-modal velocity distribution, with the majority born with spins of 50-65\% of critical rotation 
constituting the red MS, while a smaller yet significant fraction of stars born with 0-35\% of critical rotation constitute the blue MS.
This does not only agree with the bimodal velocity distribution found for field B and A type stars \citep{Dufton2013,Zorec2012}, but also matches the spectroscopically measured velocities of the red and the blue MS stars in NGC\,1818 \citep{Marino2018}. 

For our analysis, we found it most useful to focus on the unevolved stars well below the MS turnoff, because near the turnoff, gravity darkening as well as single- and binary-star evolution effects can blur the picture. We show that model sets which use different assumptions about rotational mixing lead to the same results when studying the unevolved stars, but predict different turn-off features. For example, fast-rotating SYCLIST models that adopt efficient rotational mixing are bluer than their slowly-rotating counterparts in the turn-off region. In contrast, in our MESA models and in the MIST models, in which strong rotational mixing is inhibited by chemical composition gradients, fast-rotating models always remain redder than their slowly-rotating counterparts. Current sparse spectroscopic observations find that fast-rotating stars are in general redder than the slowly-rotating ones even near the cluster turnoff, which may support inefficient rotational mixing \citep{Dupree2017,Marino2018,2020MNRAS.492.2177K}.  
The surface nitrogen abundance is significantly enhanced in both, MESA and SYCLIST fast-rotating models, but the helium abundance is only enhanced in the fast-rotating SYCLIST models.
Spectroscopic measurements of the surface chemical composition of stars in the turn-off region may be able to put tighter constraints on the strength of rotational mixing.

The knowledge of the initial spin distribution of the cluster stars bears important conclusions
for their later evolution. Consistent with previous studies, our single-star models increase their critical velocity fraction as consequence of their contracting core during hydrogen burning. 
We found that if the rotation-velocity threshold for a star to be a Be star is as low as 70\% of the critical velocity, our LMC models starting with 50-65\% of their critical velocity can evolve into Be stars before the end of core hydrogen burning. However, for a threshold larger than 80\% of critical velocity, it depends on the cluster age whether the initially rapidly rotating single-star models can become Be stars during their MS evolution. A major factor that may prevent stellar models from rotating critically is the strong mass loss due to the bi-stability jump. As discussed above, whether the bi-stability jump exists is disputed. Nevertheless, the uncertain wind mass loss does not affect our finding that in most cases the initially moderately rotating stars (red MS stars) cannot reproduce the Be stars as faint as 2-3 magnitudes below the cluster turnoff.
The detection of these faint Be stars in young star clusters implies that binary evolution is an indispensable part of Be star formation, while single-star evolution may still play a role in producing Be stars near the cluster turnoff.

Our conclusion that the vast majority of stars are born with moderately rapid or with slow rotation also affects our understanding of post-MS stars. It implies, that according to our models, rotational mixing may affect the surface abundances of trace elements like boron or nitrogen. But otherwise, rotation is not affecting the evolution of the stars, their post-MS life, supernova type, or the remnant mass in an important way. 
We might still expect differences in the advanced evolution of blue and red MS stars, perhaps due to different binary properties or different magnetic fields, which would reward study with appropriate stellar models and investigation in future observational campaigns.

\begin{acknowledgements}
We thank Dominic Bowman for useful discussions on how asteroseismology constrains the rotational structure of stars. We thank Sebastian Kamann and Nate Bastian for their stimulating discussions. SdM and SJ acknowledge partial funding by the Netherlands Organization for Scientific Research (NWO) as part of the Vidi research program BinWaves (project number 639.042.728). This work has received funding from the European Research Council (ERC) under the European Union's Horizon 2020 research innovation programme (Grant Agreement ERC-StG 2016, No 716082 'GALFOR', PI: Milone, http://progetti.dfa.unipd.it/GALFOR), and from MIUR through the FARE project R164RM93XW SEMPLICE (PI: Milone) and the PRIN program 2017Z2HSMF (PI: Bedin).
\end{acknowledgements}

\begin{appendix} 
\section{Examples of stellar evolution in the HRD}\label{sec:Appendix_A}
\setcounter{figure}{0}
\renewcommand\thesection{\Alph{section}}
\renewcommand{\thefigure}{\thesection.\arabic{figure}}
Figure\,\ref{fig:HRD_5} displays the evolutionary tracks of 5\mso models with different initial rotational velocities in the HRD. The left and the right panels correspond to the LMC and SMC metallicity, respectively. We take into account our MESA models, the SYCLIST models and the MIST models. We have redefined the initial critical velocity fraction for the SYCLIST models, such that it matches the definition used in our MESA models (see Sec.\,\ref{sec:method}). As a reference, we show the formally labeled critical angular velocity fraction of the SYCLIST models in brackets. As mentioned in Sec.\,\ref{sec:method}, the SYCLIST models use a larger metallicity $Z$ appropriate for the LMC stars than our MESA models and the MIST models. 
As a consequence, the SYCLIST ZAMS models are cooler than the corresponding MESA models in this work and the MIST models at the same velocities. Whereas all three model sets adopt similar metallicity tailored for the SMC stars, therefore, their ZAMS models occupy similar positions in the HRD.

It can be seen in Fig.\,\ref{fig:HRD_5} that at the ZAMS, when evolutionary effects have not kicked in, the fast-rotating models are cooler than the slowly-rotating models in all three model sets, with similar rotation difference resulting in similar temperature variation. This means that the implementation of the impact of centrifugal force in the three model sets is consistent. 
Nevertheless, after the ZAMS, the models in different model sets evolve differently, even for the nonrotating-star models. The discrepancy of the evolutionary tracks of the nonrotating stars is caused by different internal mixing. For a 5\mso star, $\alpha_{\rm OV}$ roughly equals 0.1, 0.15 and 0.2 in terms of step overshooting in the SYCLIST models, our MESA models and the MIST models, respectively.
Consequently, the SYCLIST nonrotating-star models have the lowest turn-off luminosities and the shortest MS lifetimes, while the MIST models are opposite. 

\begin{figure} 
	\includegraphics[width=\linewidth]{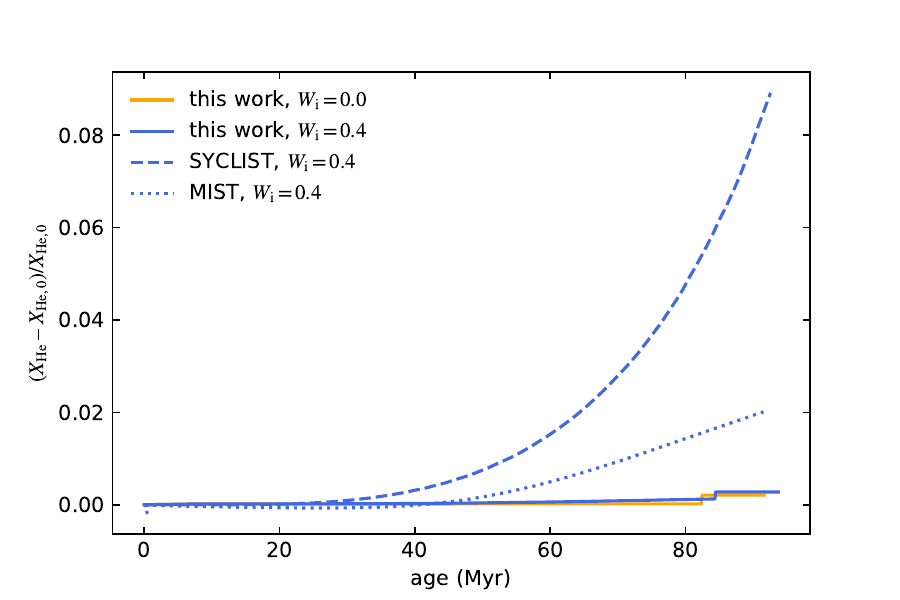}

	\centering
	\caption{Surface helium enhancement of 5\mso SMC stellar models as a function of stellar age. The y-axis shows the increase of the surface helium abundance of the stellar models with respect to their initial abundance. Here $X_\mathrm{He}$ and $X_\mathrm{He,0}$ are the current surface helium abundance and the initial surface helium abundance of the stellar models, respectively. The orange solid line corresponds to our nonrotating MESA models, while the solid, dashed, and dotted blue lines correspond to our MESA models, the SYCLIST models, and the MIST models that initially rotate at 40\% of their critical velocities.}
	\label{fig:N_5Msun} 
\end{figure}

Whereas the diverse evolution of the rotating-star models in different model sets is mainly due to different rotationally induced mixing efficiencies.
In contrast to our MESA models and the MIST models, in which the fast-rotating models are always cooler than their slowly-rotating counterparts, the evolutionary tracks of the SYCLIST fast-rotating-star models cross that of the nonrotating models near the end of the MS evolution. This is because in the SYCLIST models, helium is efficiently mixed into the stellar surface even in the presence of a chemical composition gradient. 
Figure\,\ref{fig:N_5Msun} compares relative enhancement of the surface He abundance of different stellar models. It can be seen that rotational mixing is the strongest in the SYCLIST models, but modest in the MIST models and negligible in our MESA models.
It is worth noting that rotational mixing increases the MS lifetime of the SYCLIST rotating models by an overall amount of $\sim$ 25\% \citep{2012A&A...537A.146E}, while it only increases the MS lifetime of our MESA and the MIST rotating models modestly. 

\begin{figure*} 
	\includegraphics[width=0.5\linewidth]{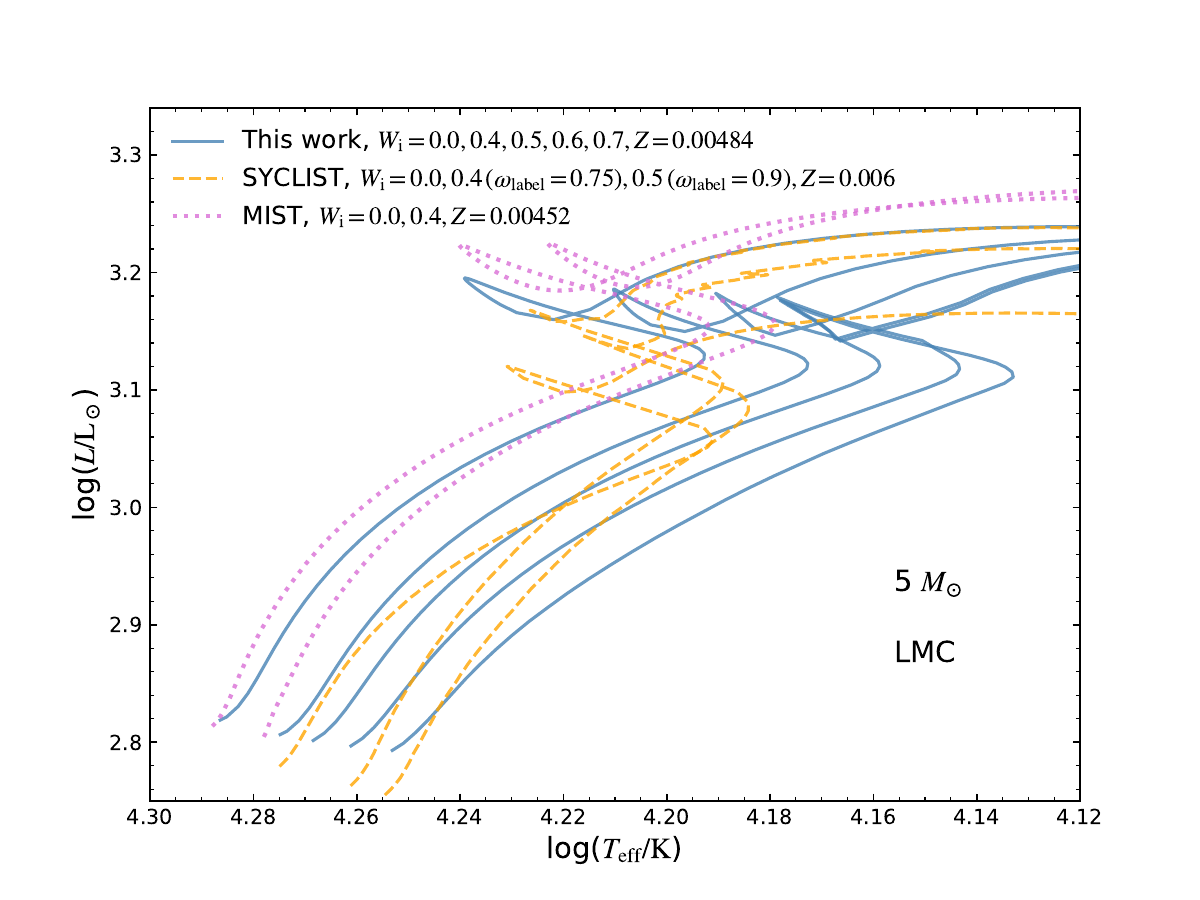}
	\includegraphics[width=0.5\linewidth]{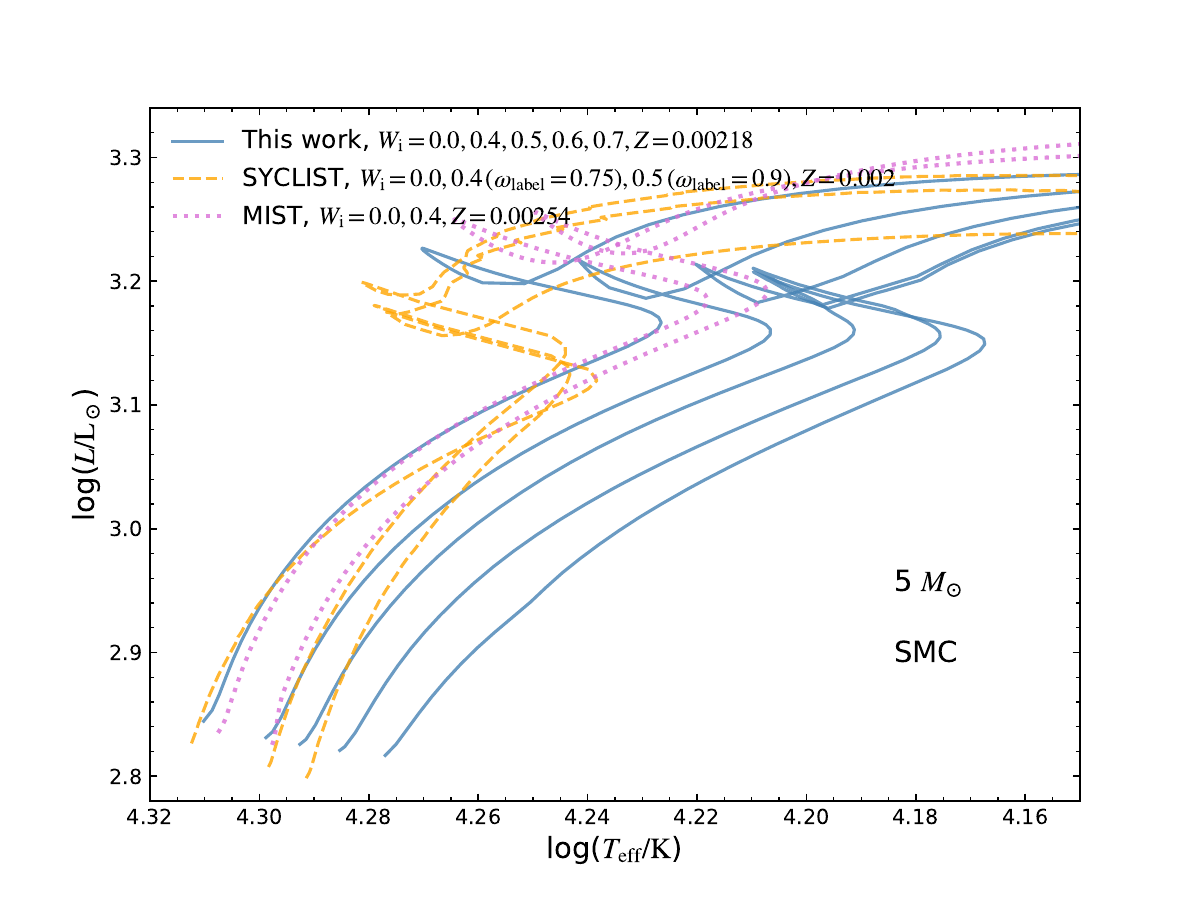}

	\centering
	\caption{Evolutionary tracks of 5\mso stellar models with LMC-like (left) and SMC-like (right) metallicities. The solid blue, dashed orange and dotted purple lines denote our MESA models, the SYCLIST models, and the MIST models, respectively. The adopted metallicity Z in different model sets are listed in the legend. In each model set, stellar evolutionary tracks from left to right correspond to models with increasing initial rotational velocity, with values listed in the legend. Numbers in the parentheses indicate the formally labeled fractional critical angular velocities in the SCYLIST web interface.}
	\label{fig:HRD_5} 
\end{figure*}

\section{Degeneracy in explaining the split main sequence as the effect of rotation}\label{sec:Appendix_B}
\setcounter{figure}{0}

\begin{figure*}[htbp]
\centering
\includegraphics[width=3.2 in]{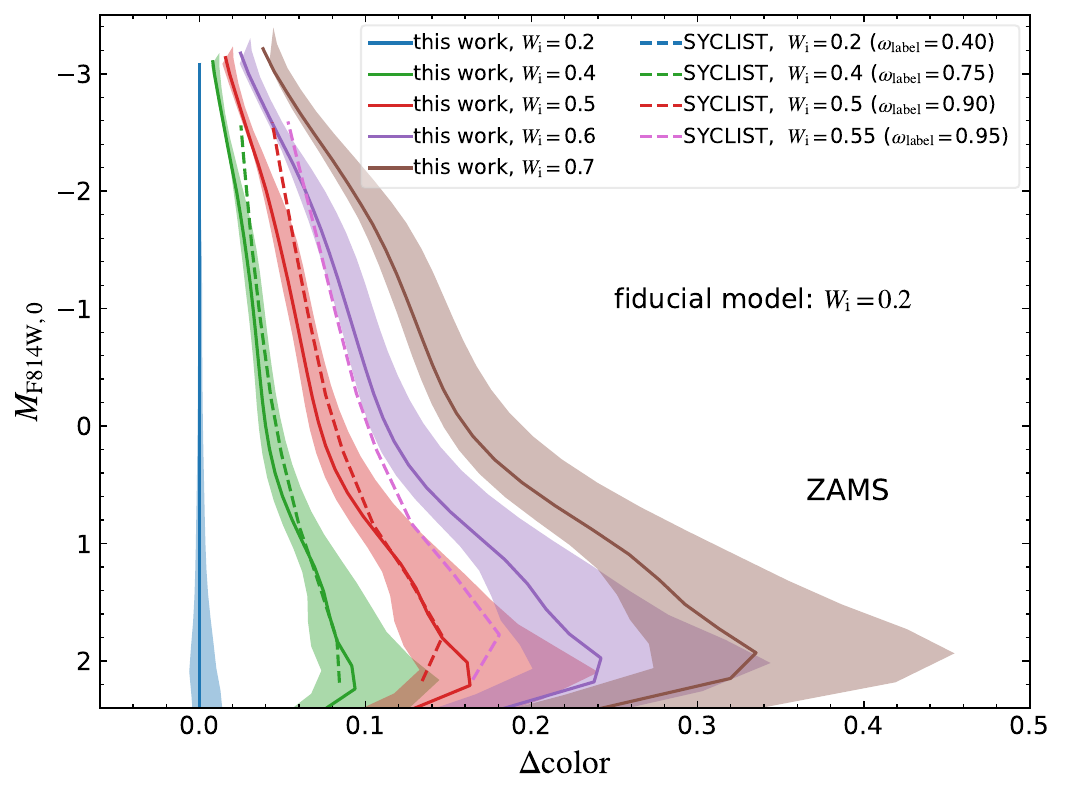}
\vspace{0.01cm}
\includegraphics[width=3.2 in]{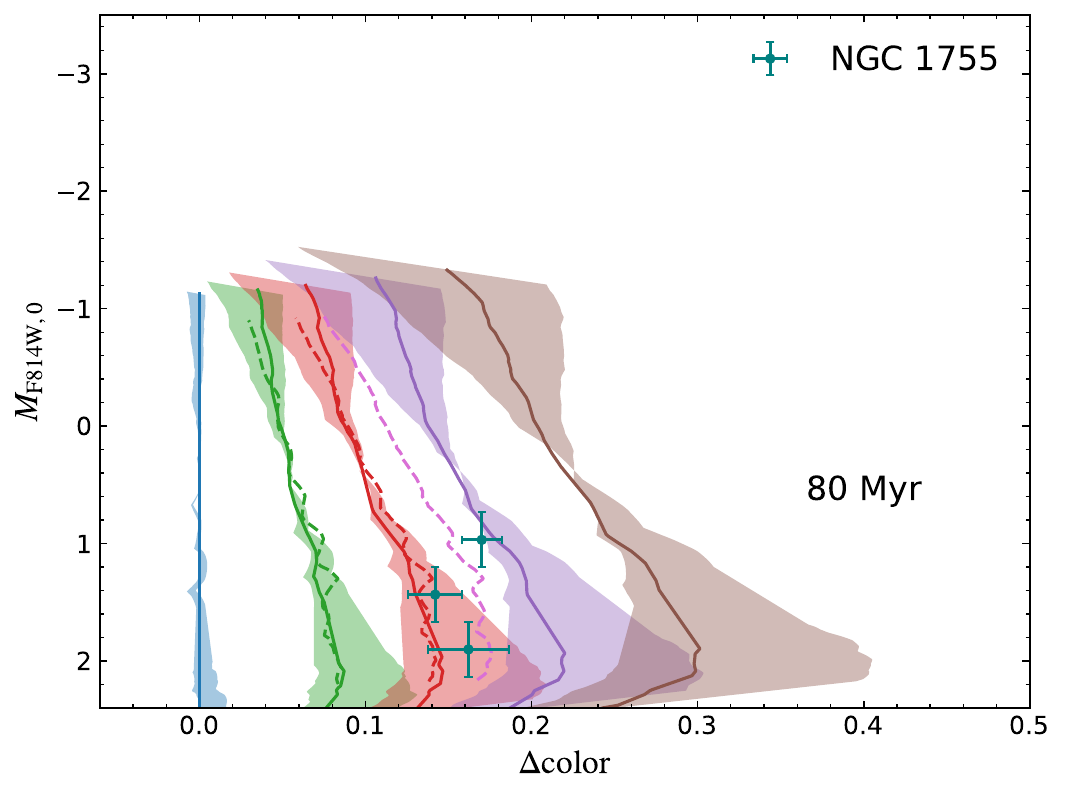}
\includegraphics[width=3.2 in]{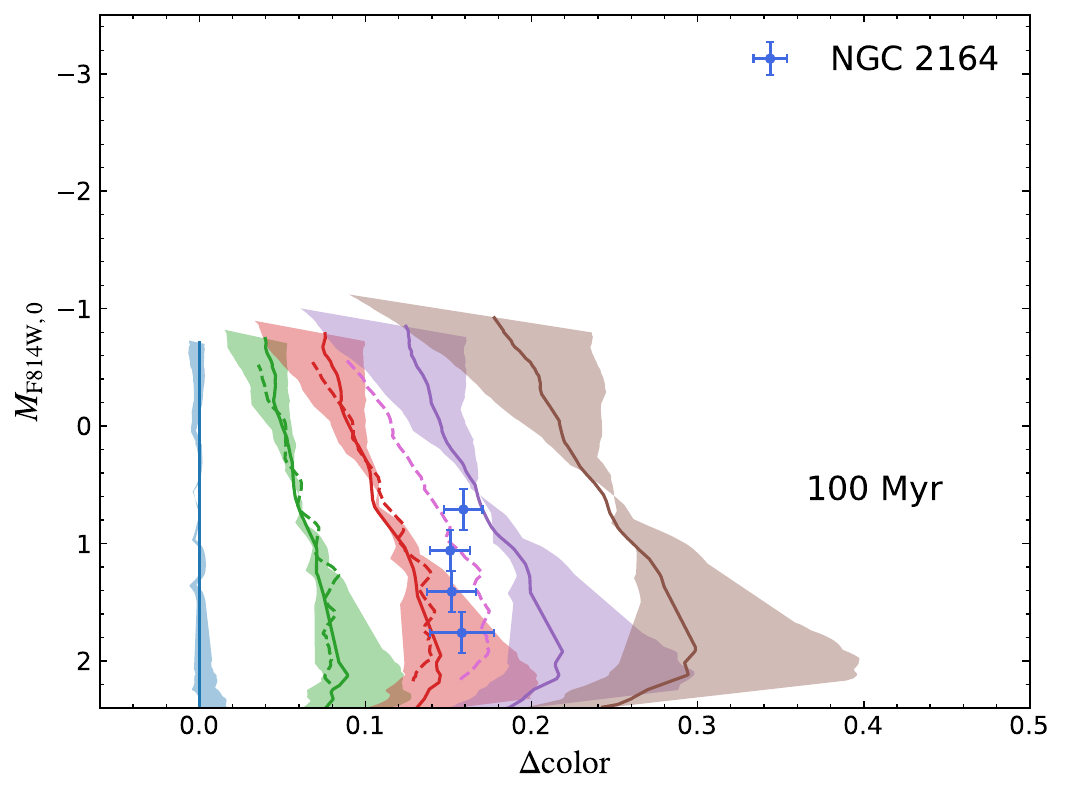}
\vspace{0.01cm}
\includegraphics[width=3.2 in]{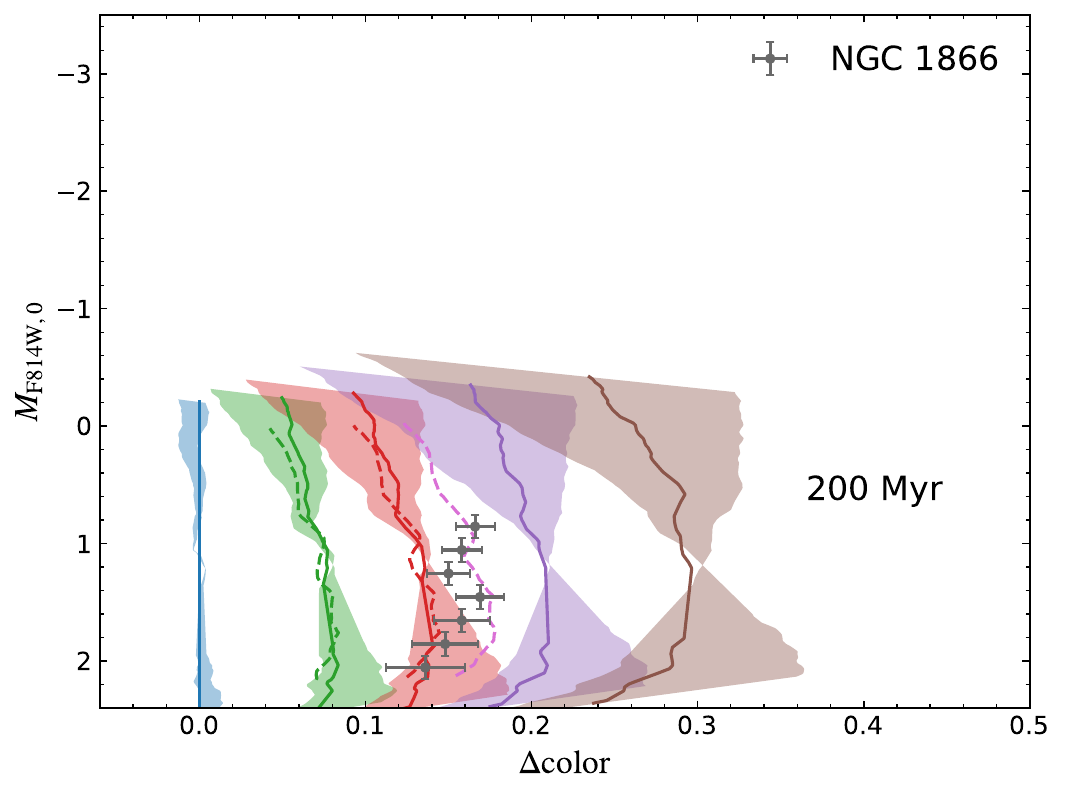}
\centering
\caption{Same as Fig.\,\ref{fig:color_split}, but color differences are calculated with respect to the $W_{\rm i}=0.2$ models.
}
 \label{fig:app_b1} 
\end{figure*}

\begin{figure*}[htbp]
\centering
\includegraphics[width=3.2 in]{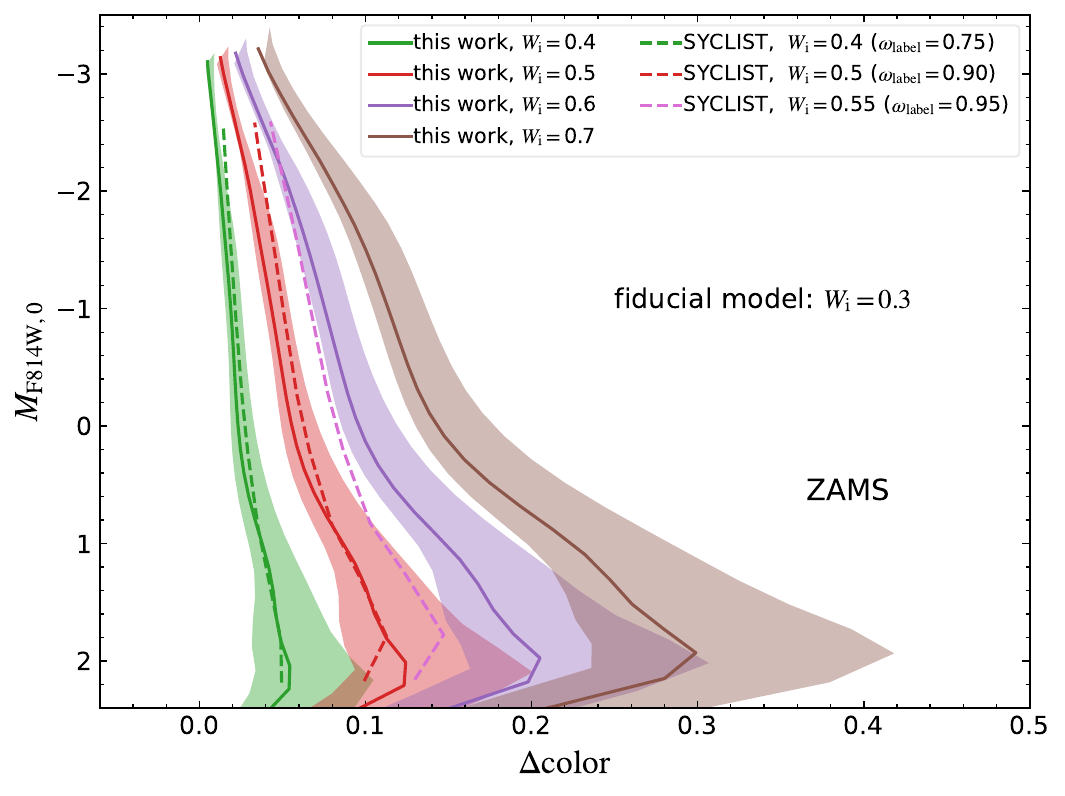}
\vspace{0.01cm}
\includegraphics[width=3.2 in]{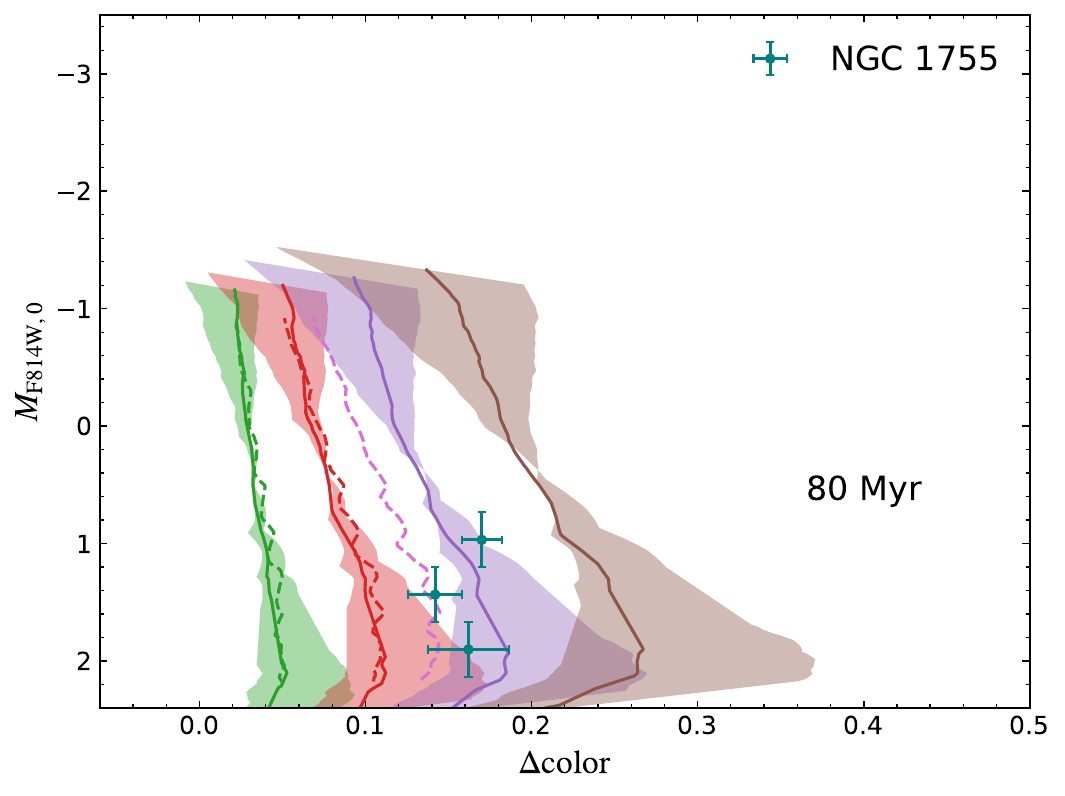}
\includegraphics[width=3.2 in]{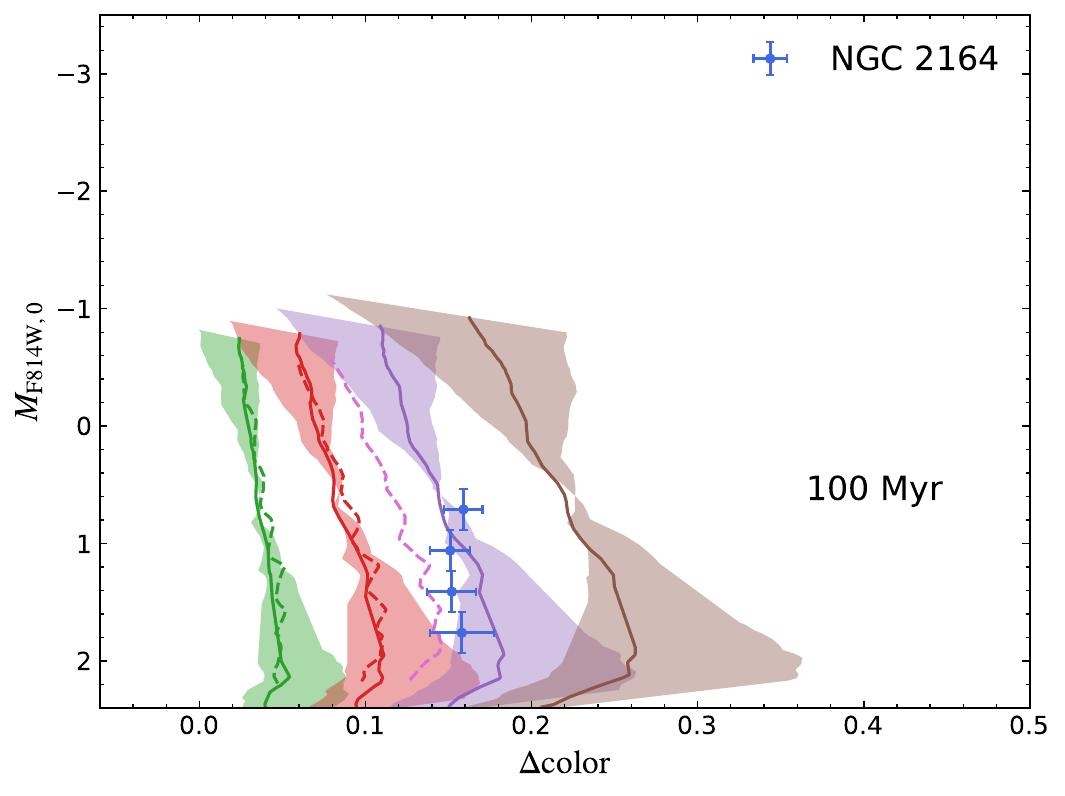}
\vspace{0.01cm}
\includegraphics[width=3.2 in]{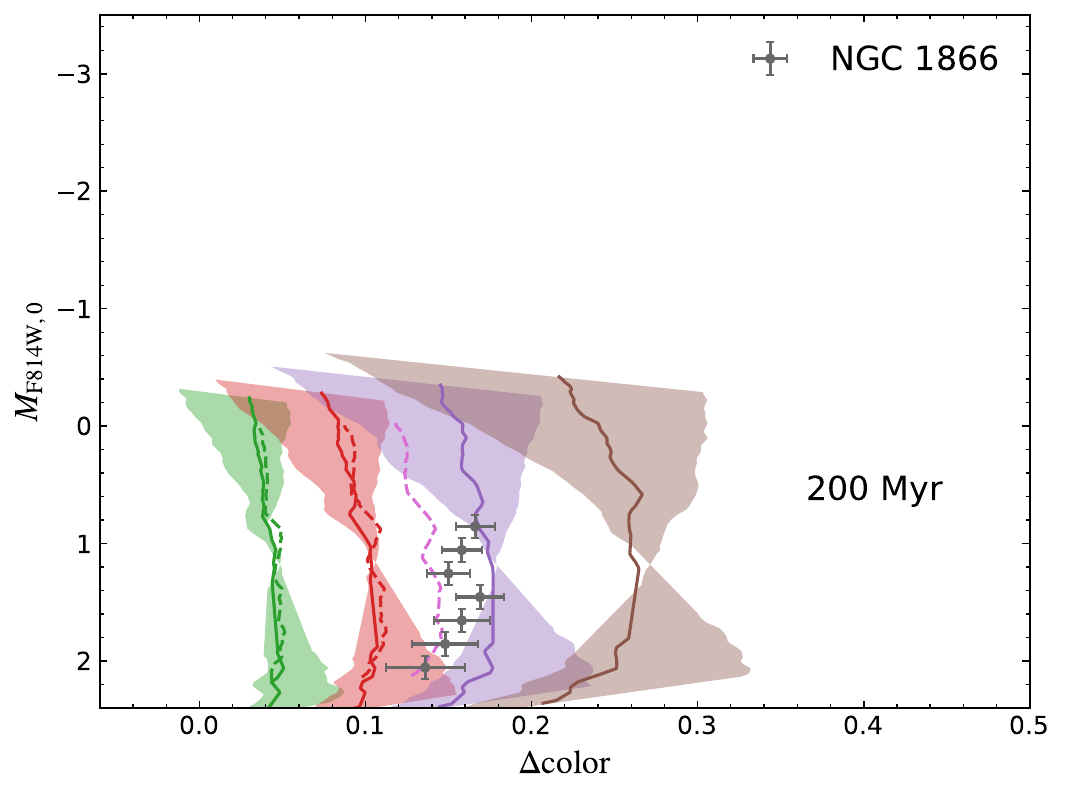}
\centering
\caption{Same as Fig.\,\ref{fig:color_split}, but color differences are calculated with respect to the $W_{\rm i}=0.3$ models.
}
 \label{fig:app_b2} 
\end{figure*}

Degeneracy exits in interpreting the observed split MS in young star clusters as the effect of rotation, because different combinations of slow and fast rotation can retrieve the observed color separation between the red and blue MSs equally well.  
Figures.\,\ref{fig:app_b1} and \ref{fig:app_b2} are similar to Fig.\,\ref{fig:color_split}, but $\Delta \mathrm{color}$ is calculated by using the isochrones of the $W_{\rm i}=0.20$, $W_{\rm i}=0.30$ star modes as the baselines. 
In these three cases, $W_{\rm i} \sim 0.55$, $W_{\rm i} \sim 0.60$ are required to explain the observed color separation between the red and blue MSs, respectively. \cite{2022NatAs...6..480W} also use $W_{\rm i} \sim 0.65$ and $W_{\rm i} \sim 0.35$ to fit the red and blue MSs in young star clusters.

\begin{figure*} 
	\includegraphics[width=0.98\linewidth]{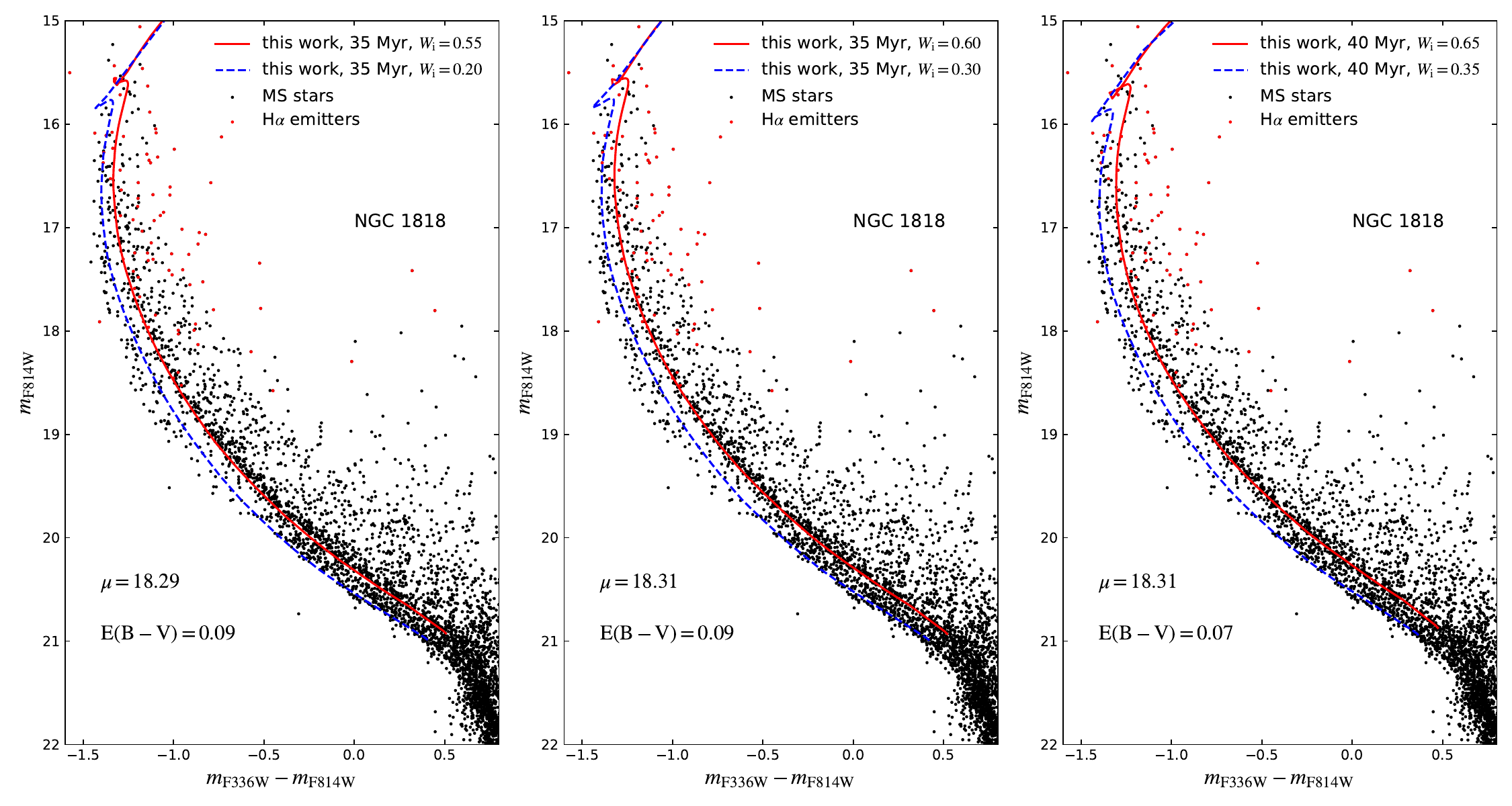}
	\centering
	\caption{Same as Fig.\,\ref{fig:iso_fit}, but we only show our MESA models and use pairs of isochrones of the $W_{\rm i}=0.55$ and $W_{\rm i}=0.20$ (left), the $W_{\rm i}=0.60$ and $W_{\rm i}=0.30$ (middle), and the $W_{\rm i}=0.65$ and $W_{\rm i}=0.35$ (right) models to fit the observed red and blue main sequences in NGC\,1818. }
	\label{fig:app_b3} 
\end{figure*}

In Fig.\,\ref{fig:app_b3}, we perform isochrone fittings to the red and the blue MS stars in NGC\,1818, using our MESA models with the abovementioned initial velocities. 
After adjusting distance modulus and reddening, we attain isochrone fittings that are as good as those shown in the right panel of Fig.\,\ref{fig:iso_fit}, which implies that photometric observations are not enough to constrain stars' precise rotation. Further spectroscopic observations are encouraged.

\section{Surface velocity of the main-sequence models}\label{sec:Appendix_D}
\setcounter{figure}{0}

\begin{figure*}[htbp] 
	\includegraphics[width=1\linewidth]{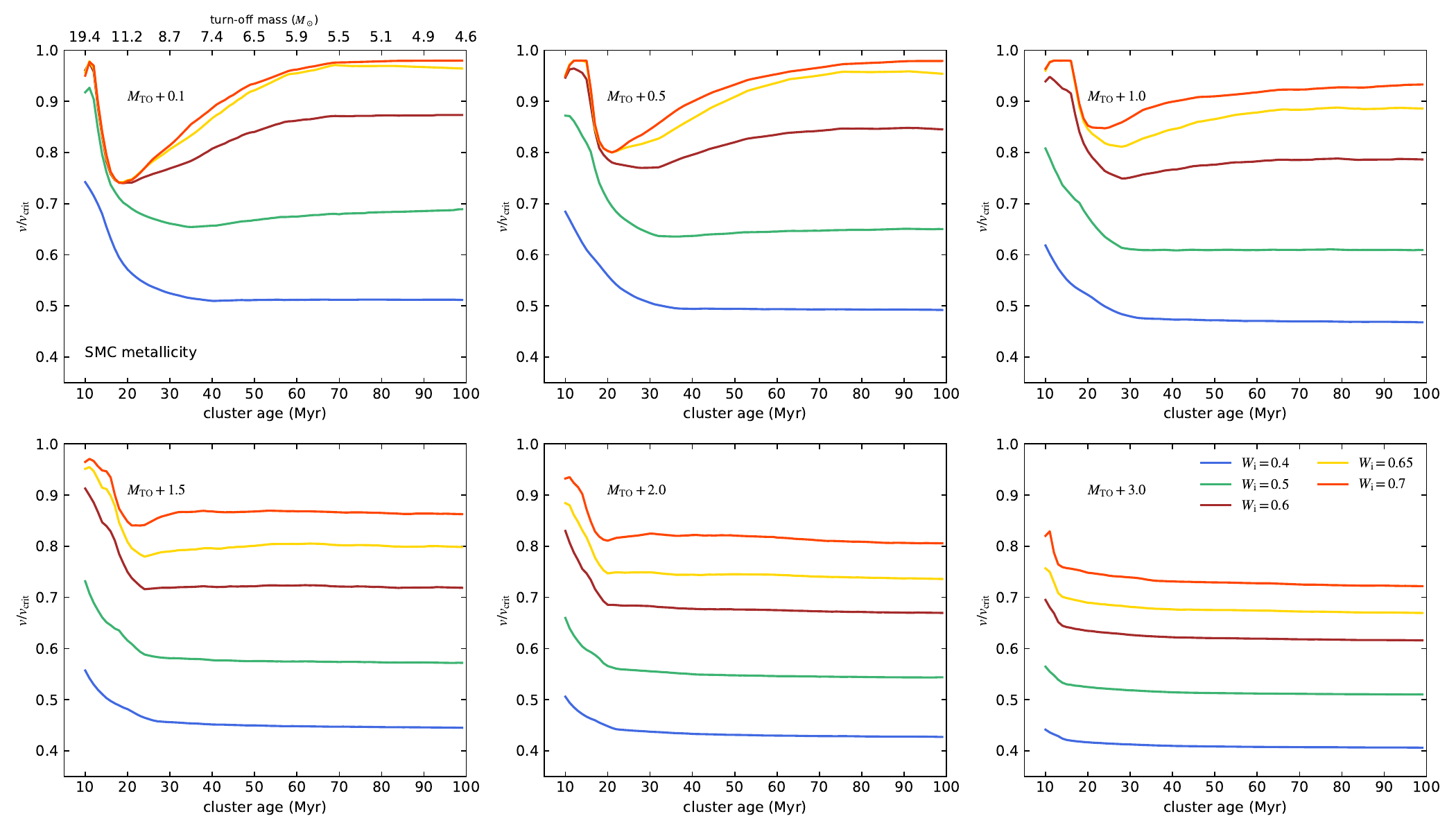}
	\centering
	\caption{Similar as Fig.\,\ref{fig:v_vcrit_age}, but for the SMC models.}
	\label{fig:app_d1} 
\end{figure*}

\begin{figure*}[htbp] 
	\includegraphics[width=0.98\linewidth]{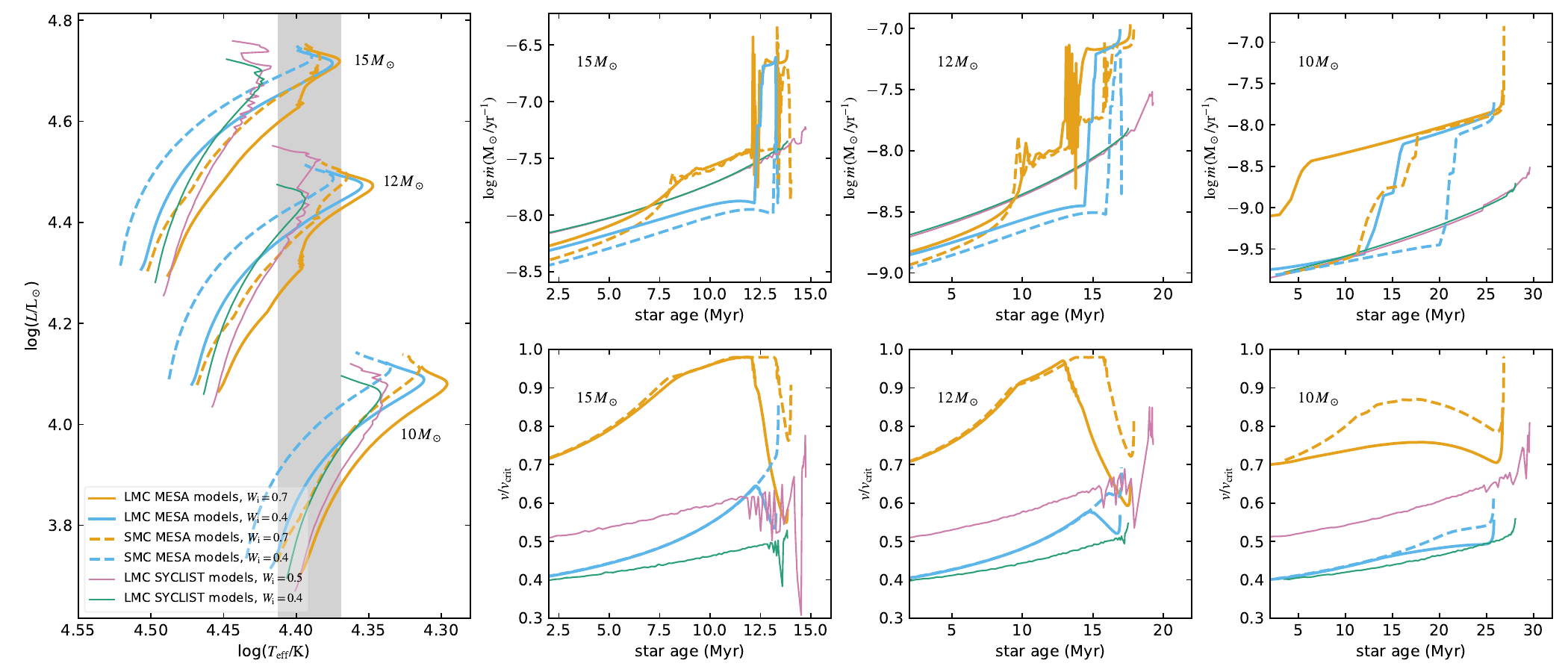}
	\centering
	\caption{The left panel shows the evolutionary tracks of the rotating single-star models in different model sets in the Hertzsprung-Russell diagram. The thick orange lines correspond to our MESA models with higher initial rotational velocity, with solid and dashed lines denoting the LMC metallicity and the SMC metallicity, respectively. While the thick blue lines represent our MESA models with lower initial rotational velocity. The thin purple and green lines denote the SYCLIST models that have initial fast and slow spins, respectively. The mass of the stellar models are listed. The gray shaded area indicates the temperature region in which we consider the bi-stability jump related stellar wind for our MESA models. The other three columns of panels illustrate the evolution of mass loss rate and critical velocity (see Eq.\ref{eq:3}) fraction as a function of stellar age, for three different masses. The meaning of the lines are the same as in the left panel.  }
	\label{fig:app_d2} 
\end{figure*}

In this appendix, we continue our discussion in Sec.\,\ref{sec:be stars} on how will our finding, that the majority of the stars should be born with moderate rotation, change our perception of Be star formation. First of all, we display Fig.\,\ref{fig:app_d1}, which is similar to Fig.\,\ref{fig:v_vcrit_age}, but for the SMC models. 
In the following, we explain more about this figure and Fig\,\ref{fig:v_vcrit_age} by showing some evolutionary examples of our MESA models and the SYCLIST models. The left panel of Fig.\,\ref{fig:app_d2} shows the evolutionary tracks of the selected stellar models in the HRD. The gray shaded area indicates the temperature zone where the stellar mass loss rate in our MESA models is affected by the bi-stability jump. The bi-stability jump temperature is set by eqs.\,14 and 15 in \citet{Vink2001}. The gray shaded area is a buffer zone with a temperature range of 10\% of the bi-stability jump temperature. In this buffer zone, the mass loss rate is computed by interpolating between the winds for hot stars \citep{Vink2001} and the winds for cool stars (i.e., the maximum value between the mass loss rate computed from \citet{Vink2001} and \citet{1990A&A...231..134N}). The right panels of this figure show the mass loss history and the surface velocity evolution of the selected stellar models. We see that our $15\,\msun$ LMC model with $W_\mathrm{i}=0.7$ achieves near critical-rotation at $\sim 12\,$Myr and then suddenly spins down to $v/v_\mathrm{crit} \sim 0.55$, which roughly equals the spin of our LMC model with $W_\mathrm{i}=0.4$ at the TAMS, due to strong stellar winds. The first peak of the mass loss rate of this star model is caused by critical rotation, while the later steady increase and the plateau between $\sim 12.5$ and 13.75\,Myr is due to the bi-stability jump. The bi-stability jump related stellar winds have a larger impact on the stellar models with higher initial velocities. Nevertheless, the $15\,\msun$ SMC star model with $W_\mathrm{i}=0.7$ does not spin down as much as its LMC counterpart because it enters the bi-stability jump temperature zone later and therefore spends less time with the enhanced stellar wind mass loss. 
A similar feature of a velocity bump before the end of MS evolution is also seen in the $15\,\msun$ models in \citet{2020A&A...633A.165H}.

The behavior of the $12\,\msun$ models is similar to the $15\,\msun$ models, but the TAMS spin of the $12\,\msun$ LMC model with $W_\mathrm{i}=0.7$ is slightly larger than that of the $15\,\msun$ counterpart, due to less intense stellar winds.
Differently, the $10\,\msun$ models reach the bi-stability jump temperature at an early time. The enhanced stellar winds prevent our MESA models from rotating critically, but they are not sufficient to significantly brake the stellar models. 
Again in this case, the SMC models achieve the bi-stability jump temperature later, and therefore, have higher surface velocities during their MS evolution than their LMC counterparts. Stars less massive than $10\,\msun$ start to get rid of the effect of the bi-stability jump and can reach near critical rotation at their TAMS, as their wind mass loss rates are quite low even in the bi-stability jump regime.
Figure\,\ref{fig:app_d2} helps us understand why the initially fast-rotating-star models in certain mass ranges cannot reach as high velocity as in other mass ranges, which is the feature seen in Fig.\,\ref{fig:v_vcrit_age}.
In contrast to our MESA models, all the selected SYCLIST models do not experience such a spin down, because they adopt \citet{1988A&AS...72..259D} winds that do not include the bi-stability jump.
We do not attempt to assess the mass loss recipes in different model sets because stellar wind is one of the least constrained physics in astronomy, but we mean to demonstrate that initially moderately rotating single stars are able to reach higher velocities, depending on the competition between angular momentum transport from the stellar center to the stellar surface and angular momentum lost accompanied by stellar winds.

\begin{figure*}[htbp] 
	\includegraphics[width=0.98\linewidth]{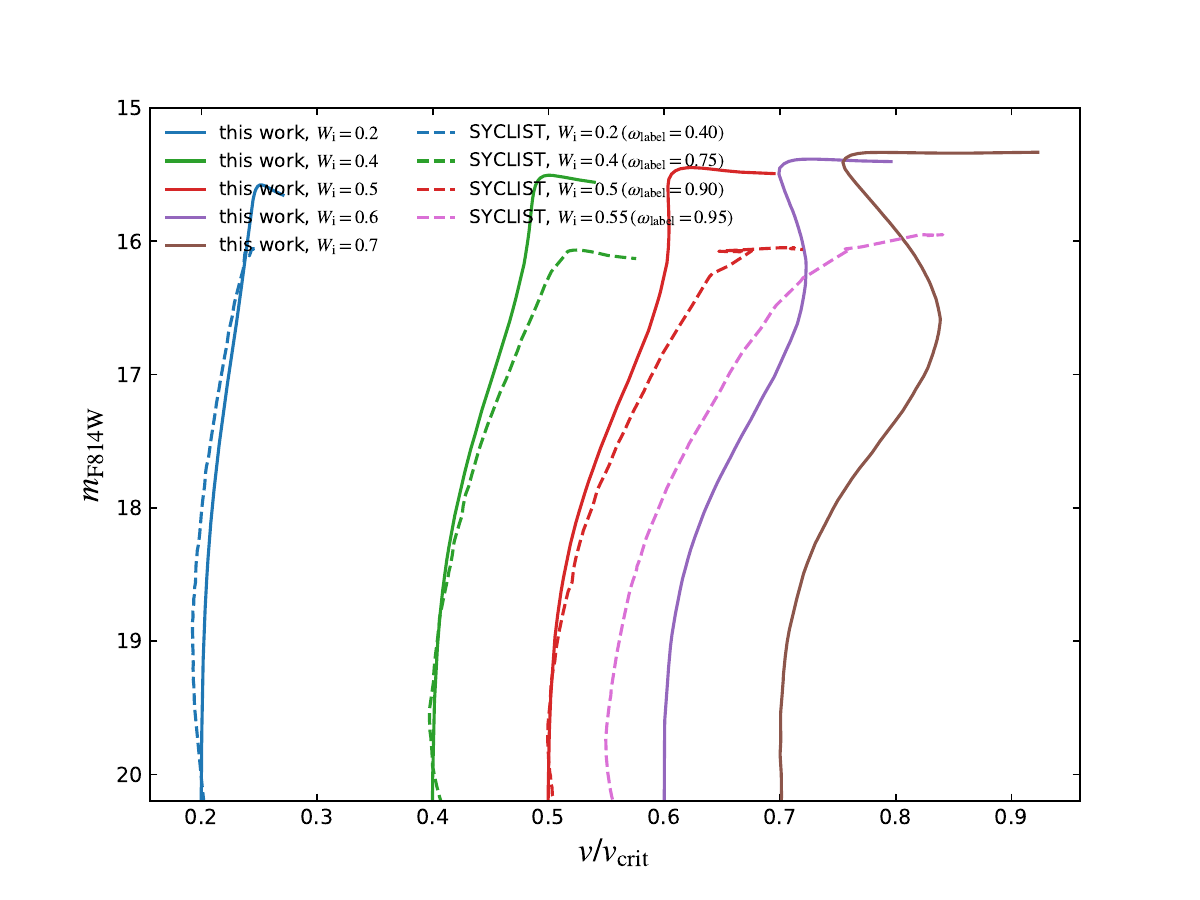}
	\centering
	\caption{Current surface rotational rates for our MESA models (solid line) and the SYCLIST models (dashed line). The adopted model age, distance modulus and reddening for each model set are the same as the right panel of Fig.\,\ref{fig:iso_fit}. Color coding is the same as Fig.\,\ref{fig:color_split}.}
	\label{fig:app_d3} 
\end{figure*}

At last, we directly examine the current surface rotational velocity of the stellar models to explore if single star evolution can interpret the observed Be stars in NGC\,1818. 
We have exhibited current velocity of the stellar models in the lower right panel of Fig.\,\ref{fig:1818_vrot}, but it is in terms of $v\sin i$ and only for the models $\sim 1.5$ magnitudes fainter than the cluster turnoff. 
Here we show current critical rotation fraction of the stellar models as a function of their apparent magnitude in Fig\,\ref{fig:app_d3}. The red solid and dashed lines are the stellar models that are employed in fitting the red MS stars in NGC\,1818 (i.e., the stellar models that are used in constructing the red and orange solid lines in the right panel of Fig.\,\ref{fig:iso_fit}).
The other lines represent the results for the stellar models with other initial velocities, using the same distance moduli and reddenings.
In general, $v/v_{\rm crit}$ increases steadily from high magnitude (unevolved part) to low magnitude (evolved part) until the turn-off magnitude, at which $v/v_{\rm crit}$ increases dramatically. 
The turn-over at around 16.5th magnitude for our MESA models with $W_\mathrm{i}\geq 0.6$ is caused by the spin-down of the turn-off stars caused by the bi-stability jump. Figure \,\ref{fig:app_d3} directly shows that the stars $\sim$1.5 magnitude fainter than the cluster turn-off magnitude almost retain their initial spins. 
It can be seen that even for the initially rapidly rotating stars ($\sim$60\% of critical velocities), single star evolution can only contribute to the Be stars at most 1.5 magnitude below the cluster turnoff, given a Be star velocity threshold of 70\% of break-up values. To explain the origin of the fainter Be stars, binary interaction is needed.

\end{appendix}


\begin{thebibliography}{78}
\expandafter\ifx\csname natexlab\endcsname\relax\def\natexlab#1{#1}\fi

\bibitem[{{Aerts}(2021)}]{2021RvMP...93a5001A}
{Aerts}, C. 2021, Reviews of Modern Physics, 93, 015001

\bibitem[{{Bastian} {et~al.}(2017){Bastian}, {Cabrera-Ziri}, {Niederhofer}, {de
  Mink}, {Georgy}, {Baade}, {Correnti}, {Usher}, \& {Romaniello}}]{Bastian2017}
{Bastian}, N., {Cabrera-Ziri}, I., {Niederhofer}, F., {et~al.} 2017, \mnras,
  465, 4795

\bibitem[{{Bastian} \& {de Mink}(2009)}]{Bastian2009}
{Bastian}, N. \& {de Mink}, S.~E. 2009, \mnras, 398, L11

\bibitem[{{Bastian} {et~al.}(2018){Bastian}, {Kamann}, {Cabrera-Ziri},
  {Georgy}, {Ekstr{\"o}m}, {Charbonnel}, {de Juan Ovelar}, \&
  {Usher}}]{Bastian2018b}
{Bastian}, N., {Kamann}, S., {Cabrera-Ziri}, I., {et~al.} 2018, \mnras, 480,
  3739

\bibitem[{{Bj{\"o}rklund} {et~al.}(2022){Bj{\"o}rklund}, {Sundqvist}, {Singh},
  {Puls}, \& {Najarro}}]{2022arXiv220308218B}
{Bj{\"o}rklund}, R., {Sundqvist}, J.~O., {Singh}, S.~M., {Puls}, J., \&
  {Najarro}, F. 2022, arXiv e-prints, arXiv:2203.08218

\bibitem[{{Bodensteiner} {et~al.}(2021){Bodensteiner}, {Sana}, {Wang},
  {Langer}, {Mahy}, {Banyard}, {de Koter}, {de Mink}, {Evans}, {G{\"o}tberg},
  {Patrick}, {Schneider}, \& {Tramper}}]{2021A&A...652A..70B}
{Bodensteiner}, J., {Sana}, H., {Wang}, C., {et~al.} 2021, \aap, 652, A70

\bibitem[{{Bodensteiner} {et~al.}(2020){Bodensteiner}, {Shenar}, \&
  {Sana}}]{2020A&A...641A..42B}
{Bodensteiner}, J., {Shenar}, T., \& {Sana}, H. 2020, \aap, 641, A42

\bibitem[{{Bowman}(2020)}]{2020FrASS...7...70B}
{Bowman}, D.~M. 2020, Frontiers in Astronomy and Space Sciences, 7, 70

\bibitem[{{Brandt} \& {Huang}(2015)}]{Brandt2015}
{Brandt}, T.~D. \& {Huang}, C.~X. 2015, \apj, 807, 25

\bibitem[{{Brott} {et~al.}(2011){Brott}, {de Mink}, {Cantiello}, {Langer}, {de
  Koter}, {Evans}, {Hunter}, {Trundle}, \& {Vink}}]{Brott2011}
{Brott}, I., {de Mink}, S.~E., {Cantiello}, M., {et~al.} 2011, \aap, 530, A115

\bibitem[{{Cantiello} \& {Langer}(2010)}]{Cantiello2010}
{Cantiello}, M. \& {Langer}, N. 2010, \aap, 521, A9

\bibitem[{{Carini} {et~al.}(2020){Carini}, {Biazzo}, {Brocato}, {Pulone}, \&
  {Pasquini}}]{2020AJ....159..152C}
{Carini}, R., {Biazzo}, K., {Brocato}, E., {Pulone}, L., \& {Pasquini}, L.
  2020, \aj, 159, 152

\bibitem[{{Chaboyer} \& {Zahn}(1992)}]{Chaboyer1992}
{Chaboyer}, B. \& {Zahn}, J.~P. 1992, \aap, 253, 173

\bibitem[{{Choi} {et~al.}(2016){Choi}, {Dotter}, {Conroy}, {Cantiello},
  {Paxton}, \& {Johnson}}]{Choi2016}
{Choi}, J., {Dotter}, A., {Conroy}, C., {et~al.} 2016, \apj, 823, 102

\bibitem[{{Correnti} {et~al.}(2017){Correnti}, {Goudfrooij}, {Bellini},
  {Kalirai}, \& {Puzia}}]{Correnti2017}
{Correnti}, M., {Goudfrooij}, P., {Bellini}, A., {Kalirai}, J.~S., \& {Puzia},
  T.~H. 2017, \mnras, 467, 3628

\bibitem[{{Cranmer}(2005)}]{2005ApJ...634..585C}
{Cranmer}, S.~R. 2005, \apj, 634, 585

\bibitem[{{D'Antona} {et~al.}(2015){D'Antona}, {Di Criscienzo}, {Decressin},
  {Milone}, {Vesperini}, \& {Ventura}}]{Antona2015}
{D'Antona}, F., {Di Criscienzo}, M., {Decressin}, T., {et~al.} 2015, \mnras,
  453, 2637

\bibitem[{{D'Antona} {et~al.}(2017){D'Antona}, {Milone}, {Tailo}, {Ventura},
  {Vesperini}, \& {di Criscienzo}}]{Antona2017}
{D'Antona}, F., {Milone}, A.~P., {Tailo}, M., {et~al.} 2017, Nature Astronomy,
  1, 0186

\bibitem[{{de Jager} {et~al.}(1988){de Jager}, {Nieuwenhuijzen}, \& {van der
  Hucht}}]{1988A&AS...72..259D}
{de Jager}, C., {Nieuwenhuijzen}, H., \& {van der Hucht}, K.~A. 1988, \aaps,
  72, 259

\bibitem[{{Dufton} {et~al.}(2013){Dufton}, {Langer}, {Dunstall}, {Evans},
  {Brott}, {de Mink}, {Howarth}, {Kennedy}, {McEvoy}, {Potter},
  {Ram{\'\i}rez-Agudelo}, {Sana}, {Sim{\'o}n-D{\'\i}az}, {Taylor}, \&
  {Vink}}]{Dufton2013}
{Dufton}, P.~L., {Langer}, N., {Dunstall}, P.~R., {et~al.} 2013, \aap, 550,
  A109

\bibitem[{{Dupree} {et~al.}(2017){Dupree}, {Dotter}, {Johnson}, {Marino},
  {Milone}, {Bailey}, {Crane}, {Mateo}, \& {Olszewski}}]{Dupree2017}
{Dupree}, A.~K., {Dotter}, A., {Johnson}, C.~I., {et~al.} 2017, \apjl, 846, L1

\bibitem[{{Ekstr{\"o}m} {et~al.}(2012){Ekstr{\"o}m}, {Georgy}, {Eggenberger},
  {Meynet}, {Mowlavi}, {Wyttenbach}, {Granada}, {Decressin}, {Hirschi},
  {Frischknecht}, {Charbonnel}, \& {Maeder}}]{2012A&A...537A.146E}
{Ekstr{\"o}m}, S., {Georgy}, C., {Eggenberger}, P., {et~al.} 2012, \aap, 537,
  A146

\bibitem[{{Ekstr{\"o}m} {et~al.}(2008){Ekstr{\"o}m}, {Meynet}, {Maeder}, \&
  {Barblan}}]{Ekstrom2008}
{Ekstr{\"o}m}, S., {Meynet}, G., {Maeder}, A., \& {Barblan}, F. 2008, \aap,
  478, 467

\bibitem[{{Endal} \& {Sofia}(1976)}]{Endal1976}
{Endal}, A.~S. \& {Sofia}, S. 1976, \apj, 210, 184

\bibitem[{{Endal} \& {Sofia}(1978)}]{1978ApJ...220..279E}
{Endal}, A.~S. \& {Sofia}, S. 1978, \apj, 220, 279

\bibitem[{{Evans} {et~al.}(2005){Evans}, {Smartt}, {Lee}, {Lennon}, {Kaufer},
  {Dufton}, {Trundle}, {Herrero}, {Sim{\'o}n-D{\'\i}az}, {de Koter}, {Hamann},
  {Hendry}, {Hunter}, {Irwin}, {Korn}, {Kudritzki}, {Langer}, {Mokiem},
  {Najarro}, {Pauldrach}, {Przybilla}, {Puls}, {Ryans}, {Urbaneja}, {Venn}, \&
  {Villamariz}}]{2005A&A...437..467E}
{Evans}, C.~J., {Smartt}, S.~J., {Lee}, J.~K., {et~al.} 2005, \aap, 437, 467

\bibitem[{{Gaia Collaboration} {et~al.}(2018){Gaia Collaboration}, {Helmi},
  {van Leeuwen}, {McMillan}, {Massari}, {Antoja}, {Robin}, {Lindegren},
  {Bastian}, {Arenou}, {Babusiaux}, {Biermann}, {Breddels}, {Hobbs}, {Jordi},
  {Pancino}, {Reyl{\'e}}, {Veljanoski}, {Brown}, {Vallenari}, {Prusti}, {de
  Bruijne}, {Bailer-Jones}, {Evans}, {Eyer}, {Jansen}, {Klioner}, {Lammers},
  {Luri}, {Mignard}, {Panem}, {Pourbaix}, {Randich}, {Sartoretti}, {Siddiqui},
  {Soubiran}, {Walton}, {Cropper}, {Drimmel}, {Katz}, {Lattanzi}, {Bakker},
  {Cacciari}, {Casta{\~n}eda}, {Chaoul}, {Cheek}, {De Angeli}, {Fabricius},
  {Guerra}, {Holl}, {Masana}, {Messineo}, {Mowlavi}, {Nienartowicz}, {Panuzzo},
  {Portell}, {Riello}, {Seabroke}, {Tanga}, {Th{\'e}venin}, {Gracia-Abril},
  {Comoretto}, {Garcia-Reinaldos}, {Teyssier}, {Altmann}, {Andrae}, {Audard},
  {Bellas-Velidis}, {Benson}, {Berthier}, {Blomme}, {Burgess}, {Busso},
  {Carry}, {Cellino}, {Clementini}, {Clotet}, {Creevey}, {Davidson}, {De
  Ridder}, {Delchambre}, {Dell'Oro}, {Ducourant},
  {Fern{\'a}ndez-Hern{\'a}ndez}, {Fouesneau}, {Fr{\'e}mat}, {Galluccio},
  {Garc{\'\i}a-Torres}, {Gonz{\'a}lez-N{\'u}{\~n}ez}, {Gonz{\'a}lez-Vidal},
  {Gosset}, {Guy}, {Halbwachs}, {Hambly}, {Harrison}, {Hern{\'a}ndez},
  {Hestroffer}, {Hodgkin}, {Hutton}, {Jasniewicz}, {Jean-Antoine-Piccolo},
  {Jordan}, {Korn}, {Krone-Martins}, {Lanzafame}, {Lebzelter}, {L{\"o}ffler},
  {Manteiga}, {Marrese}, {Mart{\'\i}n-Fleitas}, {Moitinho}, {Mora}, {Muinonen},
  {Osinde}, {Pauwels}, {Petit}, {Recio-Blanco}, {Richards}, {Rimoldini},
  {Sarro}, {Siopis}, {Smith}, {Sozzetti}, {S{\"u}veges}, {Torra}, {van Reeven},
  {Abbas}, {Abreu Aramburu}, {Accart}, {Aerts}, {Altavilla}, {{\'A}lvarez},
  {Alvarez}, {Alves}, {Anderson}, {Andrei}, {Anglada Varela}, {Antiche},
  {Arcay}, {Astraatmadja}, {Bach}, {Baker}, {Balaguer-N{\'u}{\~n}ez}, {Balm},
  {Barache}, {Barata}, {Barbato}, {Barblan}, {Barklem}, {Barrado}, {Barros},
  {Barstow}, {Bartholom{\'e} Mu{\~n}oz}, {Bassilana}, {Becciani}, {Bellazzini},
  {Berihuete}, {Bertone}, {Bianchi}, {Bienaym{\'e}}, {Blanco-Cuaresma}, {Boch},
  {Boeche}, {Bombrun}, {Borrachero}, {Bossini}, {Bouquillon}, {Bourda},
  {Bragaglia}, {Bramante}, {Bressan}, {Brouillet}, {Br{\"u}semeister},
  {Brugaletta}, {Bucciarelli}, {Burlacu}, {Busonero}, {Butkevich}, {Buzzi},
  {Caffau}, {Cancelliere}, {Cannizzaro}, {Cantat-Gaudin}, {Carballo},
  {Carlucci}, {Carrasco}, {Casamiquela}, {Castellani}, {Castro-Ginard},
  {Charlot}, {Chemin}, {Chiavassa}, {Cocozza}, {Costigan}, {Cowell}, {Crifo},
  {Crosta}, {Crowley}, {Cuypers}, {Dafonte}, {Damerdji}, {Dapergolas}, {David},
  {David}, {de Laverny}, {De Luise}, {De March}, {de Martino}, {de Souza}, {de
  Torres}, {Debosscher}, {del Pozo}, {Delbo}, {Delgado}, {Delgado}, {Di
  Matteo}, {Diakite}, {Diener}, {Distefano}, {Dolding}, {Drazinos},
  {Dur{\'a}n}, {Edvardsson}, {Enke}, {Eriksson}, {Esquej}, {Eynard Bontemps},
  {Fabre}, {Fabrizio}, {Faigler}, {Falc{\~a}o}, {Farr{\`a}s Casas}, {Federici},
  {Fedorets}, {Fernique}, {Figueras}, {Filippi}, {Findeisen}, {Fonti},
  {Fraile}, {Fraser}, {Fr{\'e}zouls}, {Gai}, {Galleti}, {Garabato},
  {Garc{\'\i}a-Sedano}, {Garofalo}, {Garralda}, {Gavel}, {Gavras}, {Gerssen},
  {Geyer}, {Giacobbe}, {Gilmore}, {Girona}, {Giuffrida}, {Glass}, {Gomes},
  {Granvik}, {Gueguen}, {Guerrier}, {Guiraud}, {Guti{\'e}rrez-S{\'a}nchez},
  {Hofmann}, {Holland}, {Huckle}, {Hypki}, {Icardi}, {Jan{\ss}en}, {Jevardat de
  Fombelle}, {Jonker}, {Juh{\'a}sz}, {Julbe}, {Karampelas}, {Kewley}, {Klar},
  {Kochoska}, {Kohley}, {Kolenberg}, {Kontizas}, {Kontizas}, {Koposov},
  {Kordopatis}, {Kostrzewa-Rutkowska}, {Koubsky}, {Lambert}, {Lanza}, {Lasne},
  {Lavigne}, {Le Fustec}, {Le Poncin-Lafitte}, {Lebreton}, {Leccia}, {Leclerc},
  {Lecoeur-Taibi}, {Lenhardt}, {Leroux}, {Liao}, {Licata}, {Lindstr{\o}m},
  {Lister}, {Livanou}, {Lobel}, {L{\'o}pez}, {Managau}, {Mann}, {Mantelet},
  {Marchal}, {Marchant}, {Marconi}, {Marinoni}, {Marschalk{\'o}}, {Marshall},
  {Martino}, {Marton}, {Mary}, {Matijevi{\v{c}}}, {Mazeh}, {Messina},
  {Michalik}, {Millar}, {Molina}, {Molinaro}, {Moln{\'a}r}, {Montegriffo},
  {Mor}, {Morbidelli}, {Morel}, {Morris}, {Mulone}, {Muraveva}, {Musella},
  {Nelemans}, {Nicastro}, {Noval}, {O'Mullane}, {Ord{\'e}novic},
  {Ord{\'o}{\~n}ez-Blanco}, {Osborne}, {Pagani}, {Pagano}, {Pailler},
  {Palacin}, {Palaversa}, {Panahi}, {Pawlak}, {Piersimoni}, {Pineau}, {Plachy},
  {Plum}, {Poggio}, {Poujoulet}, {Pr{\v{s}}a}, {Pulone}, {Racero}, {Ragaini},
  {Rambaux}, {Ramos-Lerate}, {Regibo}, {Riclet}, {Ripepi}, {Riva}, {Rivard},
  {Rixon}, {Roegiers}, {Roelens}, {Romero-G{\'o}mez}, {Rowell}, {Royer},
  {Ruiz-Dern}, {Sadowski}, {Sagrist{\`a} Sell{\'e}s}, {Sahlmann}, {Salgado},
  {Salguero}, {Sanna}, {Santana-Ros}, {Sarasso}, {Savietto}, {Schultheis},
  {Sciacca}, {Segol}, {Segovia}, {S{\'e}gransan}, {Shih}, {Siltala}, {Silva},
  {Smart}, {Smith}, {Solano}, {Solitro}, {Sordo}, {Soria Nieto}, {Souchay},
  {Spagna}, {Spoto}, {Stampa}, {Steele}, {Steidelm{\"u}ller}, {Stephenson},
  {Stoev}, {Suess}, {Surdej}, {Szabados}, {Szegedi-Elek}, {Tapiador}, {Taris},
  {Tauran}, {Taylor}, {Teixeira}, {Terrett}, {Teyssandier}, {Thuillot},
  {Titarenko}, {Torra Clotet}, {Turon}, {Ulla}, {Utrilla}, {Uzzi}, {Vaillant},
  {Valentini}, {Valette}, {van Elteren}, {Van Hemelryck}, {van Leeuwen},
  {Vaschetto}, {Vecchiato}, {Viala}, {Vicente}, {Vogt}, {von Essen}, {Voss},
  {Votruba}, {Voutsinas}, {Walmsley}, {Weiler}, {Wertz}, {Wevems},
  {Wyrzykowski}, {Yoldas}, {{\v{Z}}erjal}, {Ziaeepour}, {Zorec}, {Zschocke},
  {Zucker}, {Zurbach}, \& {Zwitter}}]{2018A&A...616A..12G}
{Gaia Collaboration}, {Helmi}, A., {van Leeuwen}, F., {et~al.} 2018, \aap, 616,
  A12

\bibitem[{{Georgy} {et~al.}(2013){Georgy}, {Ekstr{\"o}m}, {Granada}, {Meynet},
  {Mowlavi}, {Eggenberger}, \& {Maeder}}]{2013A&A...553A..24G}
{Georgy}, C., {Ekstr{\"o}m}, S., {Granada}, A., {et~al.} 2013, \aap, 553, A24

\bibitem[{{Girardi} {et~al.}(2011){Girardi}, {Eggenberger}, \&
  {Miglio}}]{Girardi2011}
{Girardi}, L., {Eggenberger}, P., \& {Miglio}, A. 2011, \mnras, 412, L103

\bibitem[{{Gossage} {et~al.}(2019){Gossage}, {Conroy}, {Dotter},
  {Cabrera-Ziri}, {Dolphin}, {Bastian}, {Dalcanton}, {Goudfrooij}, {Johnson},
  {Williams}, {Rosenfield}, {Kalirai}, \& {Fouesneau}}]{2019ApJ...887..199G}
{Gossage}, S., {Conroy}, C., {Dotter}, A., {et~al.} 2019, \apj, 887, 199

\bibitem[{{Hastings} {et~al.}(2021){Hastings}, {Langer}, {Wang},
  {Schootemeijer}, \& {Milone}}]{2021A&A...653A.144H}
{Hastings}, B., {Langer}, N., {Wang}, C., {Schootemeijer}, A., \& {Milone},
  A.~P. 2021, \aap, 653, A144

\bibitem[{{Hastings} {et~al.}(2020){Hastings}, {Wang}, \&
  {Langer}}]{2020A&A...633A.165H}
{Hastings}, B., {Wang}, C., \& {Langer}, N. 2020, \aap, 633, A165

\bibitem[{{Heger} {et~al.}(2000){Heger}, {Langer}, \& {Woosley}}]{Heger2000}
{Heger}, A., {Langer}, N., \& {Woosley}, S.~E. 2000, \apj, 528, 368

\bibitem[{{Huang} {et~al.}(2010){Huang}, {Gies}, \& {McSwain}}]{Huang2010}
{Huang}, W., {Gies}, D.~R., \& {McSwain}, M.~V. 2010, \apj, 722, 605

\bibitem[{{Inno} {et~al.}(2016){Inno}, {Bono}, {Matsunaga}, {Fiorentino},
  {Marconi}, {Lemasle}, {da Silva}, {Soszy{\'n}ski}, {Udalski}, {Romaniello},
  \& {Rix}}]{2016ApJ...832..176I}
{Inno}, L., {Bono}, G., {Matsunaga}, N., {et~al.} 2016, \apj, 832, 176

\bibitem[{{Joshi} \& {Panchal}(2019)}]{2019A&A...628A..51J}
{Joshi}, Y.~C. \& {Panchal}, A. 2019, \aap, 628, A51

\bibitem[{{Kamann} {et~al.}(2020){Kamann}, {Bastian}, {Gossage}, {Baade},
  {Cabrera-Ziri}, {Da Costa}, {de Mink}, {Georgy}, {Giesers}, {G{\"o}ttgens},
  {Hilker}, {Husser}, {Lardo}, {Larsen}, {Mackey}, {Martocchia}, {Mucciarelli},
  {Platais}, {Roth}, {Salaris}, {Usher}, \& {Yong}}]{2020MNRAS.492.2177K}
{Kamann}, S., {Bastian}, N., {Gossage}, S., {et~al.} 2020, \mnras, 492, 2177

\bibitem[{{Kamann} {et~al.}(2018){Kamann}, {Bastian}, {Husser}, {Martocchia},
  {Usher}, {den Brok}, {Dreizler}, {Kelz}, {Krajnovi{\'c}}, {Richard},
  {Steinmetz}, \& {Weilbacher}}]{2018MNRAS.480.1689K}
{Kamann}, S., {Bastian}, N., {Husser}, T.~O., {et~al.} 2018, \mnras, 480, 1689

\bibitem[{{Keller} \& {Bessell}(1998)}]{1998A&A...340..397K}
{Keller}, S.~C. \& {Bessell}, M.~S. 1998, \aap, 340, 397

\bibitem[{{Langer}(2012)}]{Langer2012}
{Langer}, N. 2012, \araa, 50, 107

\bibitem[{{Li} {et~al.}(2017){Li}, {de Grijs}, {Deng}, \&
  {Milone}}]{2017ApJ...844..119L}
{Li}, C., {de Grijs}, R., {Deng}, L., \& {Milone}, A.~P. 2017, \apj, 844, 119

\bibitem[{{Lin} {et~al.}(2011){Lin}, {Krumholz}, \&
  {Kratter}}]{2011MNRAS.416..580L}
{Lin}, M.-K., {Krumholz}, M.~R., \& {Kratter}, K.~M. 2011, \mnras, 416, 580

\bibitem[{{Liu} {et~al.}(2006){Liu}, {van Paradijs}, \& {van den
  Heuvel}}]{2006A&A...455.1165L}
{Liu}, Q.~Z., {van Paradijs}, J., \& {van den Heuvel}, E.~P.~J. 2006, \aap,
  455, 1165

\bibitem[{{Maeder} \& {Meynet}(2000)}]{Maeder2000}
{Maeder}, A. \& {Meynet}, G. 2000, \araa, 38, 143

\bibitem[{{Marchant} {et~al.}(2016){Marchant}, {Langer}, {Podsiadlowski},
  {Tauris}, \& {Moriya}}]{2016A&A...588A..50M}
{Marchant}, P., {Langer}, N., {Podsiadlowski}, P., {Tauris}, T.~M., \&
  {Moriya}, T.~J. 2016, \aap, 588, A50

\bibitem[{{Marino} {et~al.}(2018){Marino}, {Przybilla}, {Milone}, {Da Costa},
  {D'Antona}, {Dotter}, \& {Dupree}}]{Marino2018}
{Marino}, A.~F., {Przybilla}, N., {Milone}, A.~P., {et~al.} 2018, \aj, 156, 116

\bibitem[{{Meynet} \& {Maeder}(1997)}]{Meynet1997}
{Meynet}, G. \& {Maeder}, A. 1997, \aap, 321, 465

\bibitem[{{Milone} {et~al.}(2016){Milone}, {Marino}, {D'Antona}, {Bedin}, {Da
  Costa}, {Jerjen}, \& {Mackey}}]{Milone2016}
{Milone}, A.~P., {Marino}, A.~F., {D'Antona}, F., {et~al.} 2016, \mnras, 458,
  4368

\bibitem[{{Milone} {et~al.}(2017){Milone}, {Marino}, {D'Antona}, {Bedin},
  {Piotto}, {Jerjen}, {Anderson}, {Dotter}, {di Criscienzo}, \&
  {Lagioia}}]{Milone2017}
{Milone}, A.~P., {Marino}, A.~F., {D'Antona}, F., {et~al.} 2017, \mnras, 465,
  4363

\bibitem[{{Milone} {et~al.}(2018){Milone}, {Marino}, {Di Criscienzo},
  {D'Antona}, {Bedin}, {Da Costa}, {Piotto}, {Tailo}, {Dotter}, {Angeloni},
  {Anderson}, {Jerjen}, {Li}, {Dupree}, {Granata}, {Lagioia}, {Mackey},
  {Nardiello}, \& {Vesperini}}]{Milone2018}
{Milone}, A.~P., {Marino}, A.~F., {Di Criscienzo}, M., {et~al.} 2018, \mnras,
  477, 2640

\bibitem[{{Niederhofer} {et~al.}(2015){Niederhofer}, {Georgy}, {Bastian}, \&
  {Ekstr{\"o}m}}]{Niederhofer2015}
{Niederhofer}, F., {Georgy}, C., {Bastian}, N., \& {Ekstr{\"o}m}, S. 2015,
  \mnras, 453, 2070

\bibitem[{{Nieuwenhuijzen} \& {de Jager}(1990)}]{1990A&A...231..134N}
{Nieuwenhuijzen}, H. \& {de Jager}, C. 1990, \aap, 231, 134

\bibitem[{{Paxton} {et~al.}(2011){Paxton}, {Bildsten}, {Dotter}, {Herwig},
  {Lesaffre}, \& {Timmes}}]{Paxton2011}
{Paxton}, B., {Bildsten}, L., {Dotter}, A., {et~al.} 2011, \apjs, 192, 3

\bibitem[{{Paxton} {et~al.}(2013){Paxton}, {Cantiello}, {Arras}, {Bildsten},
  {Brown}, {Dotter}, {Mankovich}, {Montgomery}, {Stello}, {Timmes}, \&
  {Townsend}}]{Paxton2013}
{Paxton}, B., {Cantiello}, M., {Arras}, P., {et~al.} 2013, \apjs, 208, 4

\bibitem[{{Paxton} {et~al.}(2015){Paxton}, {Marchant}, {Schwab}, {Bauer},
  {Bildsten}, {Cantiello}, {Dessart}, {Farmer}, {Hu}, {Langer}, {Townsend},
  {Townsley}, \& {Timmes}}]{Paxton2015}
{Paxton}, B., {Marchant}, P., {Schwab}, J., {et~al.} 2015, \apjs, 220, 15

\bibitem[{{Paxton} {et~al.}(2019){Paxton}, {Smolec}, {Schwab}, {Gautschy},
  {Bildsten}, {Cantiello}, {Dotter}, {Farmer}, {Goldberg}, {Jermyn}, {Kanbur},
  {Marchant}, {Thoul}, {Townsend}, {Wolf}, {Zhang}, \& {Timmes}}]{Paxton2019}
{Paxton}, B., {Smolec}, R., {Schwab}, J., {et~al.} 2019, \apjs, 243, 10

\bibitem[{{Peters} {et~al.}(2008){Peters}, {Gies}, {Grundstrom}, \&
  {McSwain}}]{2008ApJ...686.1280P}
{Peters}, G.~J., {Gies}, D.~R., {Grundstrom}, E.~D., \& {McSwain}, M.~V. 2008,
  \apj, 686, 1280

\bibitem[{{Pietrzy{\'n}ski} {et~al.}(2013){Pietrzy{\'n}ski}, {Graczyk},
  {Gieren}, {Thompson}, {Pilecki}, {Udalski}, {Soszy{\'n}ski}, {Koz{\l}owski},
  {Konorski}, {Suchomska}, {Bono}, {Moroni}, {Villanova}, {Nardetto},
  {Bresolin}, {Kudritzki}, {Storm}, {Gallenne}, {Smolec}, {Minniti}, {Kubiak},
  {Szyma{\'n}ski}, {Poleski}, {Wyrzykowski}, {Ulaczyk}, {Pietrukowicz},
  {G{\'o}rski}, \& {Karczmarek}}]{2013Natur.495...76P}
{Pietrzy{\'n}ski}, G., {Graczyk}, D., {Gieren}, W., {et~al.} 2013, \nat, 495,
  76

\bibitem[{{Pols} {et~al.}(1991){Pols}, {Cote}, {Waters}, \&
  {Heise}}]{1991A&A...241..419P}
{Pols}, O.~R., {Cote}, J., {Waters}, L.~B.~F.~M., \& {Heise}, J. 1991, \aap,
  241, 419

\bibitem[{{Raguzova} \& {Popov}(2005)}]{2005A&AT...24..151R}
{Raguzova}, N.~V. \& {Popov}, S.~B. 2005, Astronomical and Astrophysical
  Transactions, 24, 151

\bibitem[{{Rivinius} {et~al.}(2013){Rivinius}, {Carciofi}, \&
  {Martayan}}]{2013A&ARv..21...69R}
{Rivinius}, T., {Carciofi}, A.~C., \& {Martayan}, C. 2013, \aapr, 21, 69

\bibitem[{{Salpeter}(1955)}]{Salpeter1955}
{Salpeter}, E.~E. 1955, \apj, 121, 161

\bibitem[{{Schootemeijer} {et~al.}(2018){Schootemeijer}, {G{\"o}tberg}, {de
  Mink}, {Gies}, \& {Zapartas}}]{2018A&A...615A..30S}
{Schootemeijer}, A., {G{\"o}tberg}, Y., {de Mink}, S.~E., {Gies}, D., \&
  {Zapartas}, E. 2018, \aap, 615, A30

\bibitem[{{Schootemeijer} {et~al.}(2019){Schootemeijer}, {Langer}, {Grin}, \&
  {Wang}}]{2019A&A...625A.132S}
{Schootemeijer}, A., {Langer}, N., {Grin}, N.~J., \& {Wang}, C. 2019, \aap,
  625, A132

\bibitem[{{Shao} \& {Li}(2014)}]{2014ApJ...796...37S}
{Shao}, Y. \& {Li}, X.-D. 2014, \apj, 796, 37

\bibitem[{{Spruit}(2002)}]{Spruit2002}
{Spruit}, H.~C. 2002, \aap, 381, 923

\bibitem[{{van Bever} \& {Vanbeveren}(1997)}]{1997A&A...322..116V}
{van Bever}, J. \& {Vanbeveren}, D. 1997, \aap, 322, 116

\bibitem[{{Vink} {et~al.}(2001){Vink}, {de Koter}, \& {Lamers}}]{Vink2001}
{Vink}, J.~S., {de Koter}, A., \& {Lamers}, H.~J.~G.~L.~M. 2001, \aap, 369, 574

\bibitem[{{von Zeipel}(1924)}]{Vonzeipel1924}
{von Zeipel}, H. 1924, \mnras, 84, 665

\bibitem[{{Wang} {et~al.}(2020){Wang}, {Langer}, {Schootemeijer}, {Castro},
  {Adscheid}, {Marchant}, \& {Hastings}}]{2020ApJ...888L..12W}
{Wang}, C., {Langer}, N., {Schootemeijer}, A., {et~al.} 2020, \apjl, 888, L12

\bibitem[{{Wang} {et~al.}(2022){Wang}, {Langer}, {Schootemeijer}, {Milone},
  {Hastings}, {Xu}, {Bodensteiner}, {Sana}, {Castro}, {Lennon}, {Marchant},
  {Koter}, \& {Mink}}]{2022NatAs...6..480W}
{Wang}, C., {Langer}, N., {Schootemeijer}, A., {et~al.} 2022, Nature Astronomy,
  6, 480

\bibitem[{{Wang} {et~al.}(2017){Wang}, {Gies}, \&
  {Peters}}]{2017ApJ...843...60W}
{Wang}, L., {Gies}, D.~R., \& {Peters}, G.~J. 2017, \apj, 843, 60

\bibitem[{{Wang} {et~al.}(2021){Wang}, {Gies}, {Peters}, {G{\"o}tberg},
  {Chojnowski}, {Lester}, \& {Howell}}]{2021AJ....161..248W}
{Wang}, L., {Gies}, D.~R., {Peters}, G.~J., {et~al.} 2021, \aj, 161, 248

\bibitem[{{Yang} {et~al.}(2013){Yang}, {Bi}, {Meng}, \& {Liu}}]{Yang2013}
{Yang}, W., {Bi}, S., {Meng}, X., \& {Liu}, Z. 2013, \apj, 776, 112

\bibitem[{{Yoon} {et~al.}(2006){Yoon}, {Langer}, \&
  {Norman}}]{2006A&A...460..199Y}
{Yoon}, S.~C., {Langer}, N., \& {Norman}, C. 2006, \aap, 460, 199

\bibitem[{{Yudin}(2001)}]{2001A&A...368..912Y}
{Yudin}, R.~V. 2001, \aap, 368, 912

\bibitem[{{Zorec} {et~al.}(2016){Zorec}, {Fr{\'e}mat}, {Domiciano de Souza},
  {Royer}, {Cidale}, {Hubert}, {Semaan}, {Martayan}, {Cochetti}, {Arias},
  {Aidelman}, \& {Stee}}]{2016A&A...595A.132Z}
{Zorec}, J., {Fr{\'e}mat}, Y., {Domiciano de Souza}, A., {et~al.} 2016, \aap,
  595, A132

\bibitem[{{Zorec} \& {Royer}(2012)}]{Zorec2012}
{Zorec}, J. \& {Royer}, F. 2012, \aap, 537, A120

\end{thebibliography}
\end{document}